\documentclass{ws-ijmpd}

\usepackage{amsmath}
\usepackage{amssymb}
\usepackage{cite}
\usepackage{graphicx}
\usepackage{array}

\newcommand{\beqs}{\begin{equation*}}
\newcommand{\beq}{\begin{equation}}

\newcommand{\eeqs}{\end{equation*}}
\newcommand{\eeq}{\end{equation}}

\newcommand{\beqas}{\begin{eqnarray*}}
\newcommand{\beqa}{\begin{eqnarray}}

\newcommand{\eeqas}{\end{eqnarray*}}
\newcommand{\eeqa}{\end{eqnarray}}

\newcommand{\eq}[2]{\begin{equation} #1 \label{#2} \end{equation}}

\newcommand{\eps}{\varepsilon}
\newcommand{\al}{\alpha}
\newcommand{\be}{\beta}

\newcommand{\de}{\delta}
\newcommand{\om}{\omega}

\newcommand{\la}{\lambda}
\newcommand{\si}{\sigma}

\newcommand{\Om}{\Omega}

\newcommand{\blist}{\begin{itemize}}

\newcommand{\elist}{\end{itemize}}

\providecommand{\href}[2]{#2}

\DeclareFontFamily{OT1}{rsfs}{}
\DeclareFontShape{OT1}{rsfs}{m}{n}{ <-7> rsfs5 <7-10> rsfs7 <10->rsfs10}{} 
\DeclareMathAlphabet{\mycal}{OT1}{rsfs}{m}{n}
\newcommand{\scri}{{\mycal I}}

\newcommand{\diff}{\extd}

\DeclareMathOperator{\extdm}{d}
\newcommand{\extd}{\extdm \!}

\begin{document}

\markboth{D. Grumiller}{Virtual black holes and the S-matrix}

\title{VIRTUAL BLACK HOLES AND THE S-MATRIX}
\author{\footnotesize D. GRUMILLER\footnote{E-mail: 
grumiller@itp.uni-leipzig.de}}

\address{Institute for Theoretical Physics\\ Leipzig University \\ Augustusplatz 10-11, D-04103 Leipzig, Germany}
\maketitle

\begin{abstract}

A brief review on virtual black holes is presented, with special emphasis on phenomenologically relevant issues like their influence on scattering or on the specific heat of (real) black holes. Regarding theoretical topics results important for (avoidance of) information loss are summarized.

\end{abstract}


\renewcommand{\thefootnote}{\arabic{footnote}}
\setcounter{footnote}{0}

\section{Introduction}

The definition of ``Virtual Black Holes'' needs two ingredients, namely ``Virtual'' and ``Black Holes''. 

One of the basic lessons of Quantum Field Theory (QFT) is the prediction of virtual particles. ``Virtual'' means, roughly speaking, that the particle is sufficiently off the mass shell. Probably the most spectacular macroscopic physical consequence of virtual particles is the Casimir effect \cite{Casimir:1948dh}: in the simplest setup with two infinitely large parallel conducting planes the latter induce boundary conditions upon the quantum fields which change the spectrum of virtual particles. Consequently, the vacuum energy in the configuration with the plates is smaller than in the configuration without, and thus an attractive force between the plates is generated. It can be measured with great accuracy and experiments coincide well with theoretical predictions (for a review cf.~\cite{Bordag:2001qi}). 
Also in scattering experiments virtual particles can mediate measurable interactions between real ones, although the former do not enter the asymptotic states by definition in contrast to the latter.  Moreover, every unstable particle is ``slightly virtual''.
It is not very clear where to draw the line between various degrees of virtuality (for instance, stable: no decay; metastable: decay width $\Gamma$ very small, sharp Breit-Wigner resonance; unstable: large decay width but still small as compared to mass $m$; almost virtual: decay width comparable to mass; virtual: far off the mass shell), but the rough definition above will be sufficient for the present work. A nice example displaying various degrees of virtuality is provided by toponium \cite{Fadin:1988fn} 
(particle properties are taken from \cite{Hagiwara:2002fs}). 
Toponium is built from a top and an anti-top. Mass and decay width of the top quark are $m_t\approx 178.0 \pm 4.3 \,GeV$ and $\Gamma_t\approx 1.5 \,GeV$, respectively; thus, $m_t\gg\Gamma_t$  and the top quark, if it existed freely, would be considered as unstable particle (but probably not as metastable one because $m_t\ggg\Gamma_t$ is not valid). For toponium bound states the relevant energy scale is the Bohr energy $E_B=m_t\alpha^2_s$ (natural units $c=\hbar=1$ are used in this work), where $\alpha_s\approx 0.11$. But since $E_B\approx \Gamma$ toponium bound states are ``almost virtual''. The decay products are unstable by themselves (in the preferred channel $W$ bosons and $b$ quarks are produced together with their anti-particles) and eventually decay into metastable and stable particles. Radiative corrections to all these processes are governed by virtual particles. While it may be a somewhat semantic issue whether virtual particles should be considered as ``real'', they are definitely ``real'' as far as their relevance to Nature is concerned.\footnote{In fact, there seems to be so much ``reality'' involved that sometimes even issues like the parton distribution of virtual photons are discussed \cite{Schuler:1996fc}.}

Regarding the second ingredient, there is no reasonable doubt that Black Holes (BHs) are real objects in Nature (see \cite{Schodel:2002py}; for a review on BH binaries cf.~e.g.~\cite{McClintock:2003gx}). As QFT tells us that for each real object a corresponding virtual one should exist there is no question as to the existence of Virtual Black Holes (VBHs). So in principle VBHs are of interest for physics. However, it is less obvious that they are of practical relevance to experiments, especially to those accessible in the near future. After all, macroscopic BHs are so massive that ``virtuality'' becomes as irrelevant as it does for stars like our sun -- the impact of virtual stars on physical experiments is negligible. While for physics of real BHs the most relevant objects are macroscopic ones, for the physics of VBHs the microscopic ones dominate. 

Besides purely experimental issues there is considerable theoretical interest concerning VBHs. After all, ``virtual'' implies, at least to some extent, the application of QFT-like methods, while ``BH'' implies that the theory to be quantized should be General Relativity or one of its generalizations. It is well known that quantization of gravity is a difficult task (for recent reviews cf.~e.g.~\cite{Carlip:2001wq}). 
Thus, conclusions drawn from the study of VBHs may lead to valuable insight into quantum gravity. In particular, there is the famous information paradox (for reviews cf.~e.g.~\cite{Preskill:1992tc}).
\footnote{The information paradox recently attracted some attention beyond the physics community due to the ``betrayal'' of one of the most prominent and persistent members of the ``Information loss party'', S.~Hawking, who appears to have joined the ``Unitarity party'' \cite{GR17}.}
So a natural question to ask is whether VBHs lead to information loss, and if they do, what are the consequences e.g.~for scattering of ordinary particles.

This paper is organized as follows: in Section \ref{se:2} two notions of VBHs are recalled, starting with Hawking's Euklidean version \cite{Hawking:1996ag} and ending with our Minkowskian one \cite{Grumiller:2000ah}. For technical reasons the Minkowskian definition is restricted to the context of 2D dilaton gravity, which contains (among many other models) the Schwarzschild BH. It is fair to say that this review naturally is biased towards the second definition.
Section \ref{se:3} is devoted to VBHs in scattering experiments and implications for the problem of information loss.
Quantum corrections to thermodynamical observables are the topic of Section \ref{se:4}. After a brief review of thermodynamics in 2D dilaton gravity the quantum corrected specific heat of the Witten BH/CGHS model is presented. Some open points are addressed in the concluding Section \ref{se:5}, which also mentions the possibility of VBHs in Loop Quantum Gravity and String Theory.

Although most of this work has the character of a review there are a few new comments and results in Sections \ref{se:3} and \ref{se:4}. In particular, the formula for the specific heat \eqref{eq:csother} to the best of my knowledge is new, as well as the related brief discussion of Hawking-Page like transitions in generic 2D dilaton gravity. Finally, it should be mentioned that in order to be able to appreciate some of the technical points in Sections \ref{se:2.2}, \ref{se:3.2} and \ref{se:4} a review on 2D dilaton gravity \cite{Grumiller:2002nm} may be recommended.

\section{Definitions of VBHs}\label{se:2}

\subsection{Hawking's Euklidean version}

Ever since John Wheeler's proposal of ``space-time foam'' 
\cite{wheelerrelativity} physicists have toyed with the idea of quantum 
induced topology fluctuations. This has 
culminated not only in spin-foam models, which are 
considered as a serious candidate for quantum gravity (cf.~e.g.~\cite{Baez:1999sr}), 
but also in Hawking's bubble approach 
of VBHs \cite{Hawking:1996ag}.

\begin{figure}
\center
\includegraphics[width=.4\linewidth]{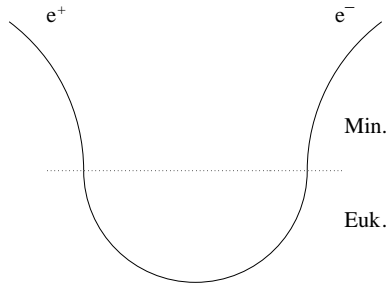}
\caption{$e^+e^-$ pair creation by tunneling through Euklidean space}
\label{fig:ep}
\end{figure}

In that paper Hawking finally abandoned the wormhole picture of spacetime foam \cite{Hawking:1988ae} 
and went back to an earlier idea, referred to as ``quantum bubble picture'' \cite{Hawking:1978zw}. So instead of regarding the first homology group as relevant, as for topologies that are multiply connected by wormholes, the second homology group was considered. The corresponding Betti number just counts the number of 2-spheres that may not be shrunk to zero (cf.~e.g.~\cite{nakaharageometry}). 
Hawking then argues that for simply connected 4-manifolds for mathematical {\em and} physical reasons only $S^2\times S^2$ has to be taken into account.\footnote{Actually, the argument is not very explicit in the original work, but it appears that Hawking invokes Wall's theorem \cite{Wall:1964} which states that after taking the connected sum with sufficiently many copies of $S^2\times S^2$ any two simply connected 4-manifolds with isomorphic intersection forms become diffeomorphic to each other. Consequently, simply connected 4-manifolds may be built by gluing (copies of) three basic units which Hawking calls ``bubbles'': projective planes ($CP^2$), Kummer-K\"ahler-Kodaira surfaces ($K3$) and VBHs ($S^2\times S^2$). Hawking dismisses $CP^2$ because it does not allow spin structure and $K3$ because it contributes to anomalies and helicity changing processes.} 
The question then arises, what has $S^2\times S^2$ to do with a VBH? It is answered by analogy to electrodynamics: in an external electric field pair creation may occur and one way to describe this process is by gluing in a sufficiently smooth way the Minkowskian solution of an electron and a positron accelerating away from each other to the Euklidean solution describing a virtual electron-positron pair (see Fig.~\ref{fig:ep}).
 
The final ingredient to Hawking's construction is the Ernst solution \cite{Ernst:1976} describing the pair creation of charged BHs in an external electric or magnetic field. By analytic continuation to the Euklidean domain one finds the topology of $S^2\times S^2$ minus a point. Analogy to electromagnetism suggests correspondence to a BH loop in a spacetime asymptotic to $\mathbb{R}^4$. Because $S^2\times S^2$ minus a point is the topological sum of $S^2\times S^2$ and $\mathbb{R}^4$ Hawking concludes that the $S^2\times S^2$ bubbles found by topological considerations summarized above may be interpreted as VBH loops. Elaborations on Euklidean VBHs include the study of loss of quantum coherence \cite{Hawking:1997ia}, of higher spin fields in VBH backgrounds \cite{Prestidge:1998bk}, of quantum evolution in spacetime foam \cite{Garay:1999cy}, of effects relevant to neutrino-oscillations \cite{Benatti:2001fa} and of non-standard Kaon-dynamics \cite{Benatti:1998vu} (the latter being inspired also by earlier work \cite{Ellis:1984jz}).

\paragraph{Physical consequences} The almost purely topological considerations above led Hawking to present several surprising physical implications \cite{Hawking:1996ag}: he argued in favor of loss of quantum coherence\footnote{Because of this, together with Hawking's U-turn regarding information loss \cite{GR17} VBHs may soon share the fate of wormholes to be abandoned by one of their creators.} (see also \cite{Banks:1984by}) 
and derived several consequences from it: the fact that the $\theta$ angle of QCD is zero, non-existence of fundamental scalar fields (and thus the prediction that the Higgs particle does not exist unless it is composed, for instance, of fermions) and the suggestion that at end of BH evaporation the Planck size remnant eventually disappears into the sea of VBHs. 
The last point is less relevant experimentally. But theoretically it appears to imply that 2D models of BH evaporation cannot describe the disappearance of BHs in a way that is non-singular. Another crucial remark is that in Hawking's picture VBHs may only be created in pairs. While a similar property holds for $e^+e^-$ -- for a good reason, namely charge conservation, which applies even to virtual particles -- it is slightly difficult to comprehend why VBHs may not be produced in singles -- after all, there does not seem to be any violation of global charges when a single BH is produced, be it real or virtual. Because it is an interesting task by itself, VBHs in Minkowski space will be considered next, where it is found that no such restriction arises. 

\subsection{VBHs in Minkowski space}\label{se:2.2}

Although clearly VBHs by the very definition of ``virtual'' do not satisfy classical field Eqs., Hawking's Euklidean definition of VBHs reminds of instantons (which are non-singular solutions of the classical field Eqs.~with finite Euklidean action, cf.~e.g.~\cite{Rajaraman:1982is}), because in both cases topology of an Euklidean configuration plays a crucial role. 
It is well-known that instantons may also be described in Minkowski space (although it is not necessarily a convenient way to describe them \cite{Rajaraman:1982is}), so naturally the question emerges whether VBHs have a Minkowskian counterpart. As shown below this turns out to be the case. For simplicity, the discussion will be restricted to 2D dilaton gravity which contains, among other models, spherically reduced gravity, i.e., the phenomenologically relevant Schwarzschild BH. Another reason to restrict to 2D is that, as Hawking and Ross state in ref.~\cite{Hawking:1997ia}, ``one can neither calculate the scattering in a general metric, nor integrate over all metrics'' (in D=4). Fortunately, in D=2 one can, as shown in the extensive work starting from the two basic papers of Kummer and Schwarz \cite{Kummer:1992bg}.

\subsubsection{A brief review of classical 2D dilaton gravity}

The purpose of this brief collection of well-known results is merely to fix the notation.\footnote{Signs of  mass $M$ and curvature scalar $r$ have been fixed conveniently such that $M>0$ for positive mass configurations and $r>0$ for dS. This is the only difference to the notations used in ref.~\cite{Grumiller:2002nm}.} For background information and refs.~the extensive review \cite{Grumiller:2002nm} may be consulted (for earlier reviews cf.~refs.~\cite{Brown:1988}). 
Thus, without further ado definitions will be listed: $e^a=e^a_\mu dx^\mu$ is the
dyad 1-form dual to $E_a$ -- i.e.\ $e^a(E_b)=\de^a_b$. Latin indices refer to an anholonomic frame, Greek indices to a holonomic one. The 1-form
$\omega$ represents the  spin-connection $\om^a{}_b=\eps^a{}_b\om$
with  the totally antisymmetric Levi-Civit{\'a} symbol $\eps_{ab}$ ($\eps_{01}=+1$). With the
flat metric $\eta_{ab}$ in light-cone coordinates
($\eta_{+-}=1=\eta_{-+}$, $\eta_{++}=0=\eta_{--}$) it reads $\eps^\pm{}_\pm=\pm 1$. The torsion 2-form is given by 
$T^\pm=(d\pm\omega)\wedge e^\pm$. The curvature 2-form $R^a{}_b$ can be represented by the 2-form $R$ defined by 
$R^a{}_b=\eps^a{}_b R$, $R=d\wedge\om$. The volume 2-form is denoted by $\epsilon = e^+\wedge e^-$. Signs and factors of the Hodge-$\ast$ operation are defined by $\ast\epsilon=1$. The quantities $\om$, $e^a$ are called ``Cartan variables''.
Since the Einstein-Hilbert action $\int_{\mathcal{M}_2}  R\propto(1-\gamma)$ yields just the Euler number for a surface with genus $\gamma$ one has to generalize it appropriately. The simplest idea is to introduce a Lagrange multiplier for curvature, $X$, also known as ``dilaton field'', and an arbitrary potential thereof, $V(X)$, in the action $\int_{\mathcal{M}_2}  \left(XR+\epsilon V(X)\right)$. 
Having introduced curvature it is natural to consider torsion as well. By analogy the first order gravity action \cite{Schaller:1994es}
\eq{
L^{\rm (1)}= \int_{\mathcal{M}_2}  \left(X_aT^a+XR+\epsilon\mathcal{V} (X^aX_a,X)\right)
}{eq:FOG}
can be motivated where $X_a$ are the Lagrange multipliers for torsion. It encompasses essentially all known dilaton theories in 2D. Actually, for most practical purposes the potential takes the form
\begin{equation}
  \label{eq:pot}
  \mathcal{V} (X^aX_a,X) = V(X) + X^+X^- U(X)\,.
\end{equation}
The action \eqref{eq:FOG} is equivalent to the frequently used second order action \cite{Russo:1992yg,Banks:1991mk}
\begin{equation}
\label{eq:GDT}
L^{(2)}=\int_{\mathcal{M}_2} d^{2}x\, \sqrt{-g}\; \left[ X \frac{-r}{2}-\frac{U(X)}{2}\; (\nabla X)^{2}+V(X)\; \right] \, ,
\end{equation}
with the same functions $U,V$ as in \eqref{eq:pot}. The curvature scalar $r$ and covariant derivative $\nabla$ are associated with the Levi-Civit\'a connection related to the metric $g_{\mu\nu}$, the determinant of which is denoted by $g$. If $\om$ is torsion-free $r\propto\ast R$. 

It is useful to introduce the following combinations of $U,V$:
\begin{equation}
  \label{eq:wI}
  I(X):=\exp{\int^X U(y)\extd y}\,,\quad w(X):=\int^XI(y)V(y)\extd y
\end{equation}
The integration constants may be absorbed, respectively, by rescalings and shifts of the mass $M$. Under dilaton dependent conformal transformations $X^a\to X^a/\Om$, $e^a\to e^a\Om$, $\om\to\om+X_ae^a\extd\,\ln{\Om}/\extd X$ Eq.~\eqref{eq:FOG} is mapped to a new action of the same type with transformed potentials $\tilde{U}$, $\tilde{V}$. Thus, it is not invariant. It turns out that only the combination $w(X)$ as defined in \eqref{eq:wI} remains invariant, so conformally invariant quantities may depend on $w$ only. Note that $I$ is positive apart from eventual boundaries (typically, $I$ may vanish in the asymptotic region and/or at singularities). It can be shown that there is always a conserved quantity ($\extd M=0$),
\begin{equation}
  \label{eq:c}
  M=-X^+X^-I(X)-w(X)\,.
\end{equation}
The classical solutions are labelled by this constant of motion. In the absence of matter there are no propagating physical degrees of freedom.

The line element in Eddington-Finkelstein gauge reads
\begin{equation}
  \label{eq:EF}
  \extd s^2 =2\diff u\,\diff \tilde{X} - 2I(X)(M+w(X))\diff u^2\,, 
\end{equation}
with $\extd\tilde{X}:=I(X)\extd X$.
Evidently there is always a Killing vector $k\cdot\partial=\partial/\partial u$ with associated Killing norm $k\cdot k=-2I(M+w)$. Since $I\neq 0$ Killing horizons are encountered at $X=X_h$ where $X_h$ is a solution of
\begin{equation}
  \label{eq:horizon}
  w(X_h)+M=0\,.
\end{equation}
In the simple conformal frame $I=1$ the curvature scalar may be expressed as
\begin{equation}
  \label{eq:curv}
  r = 2w''\,. 
\end{equation}
Note that the independence of curvature from the mass $M$ and from $I(X)$ is an artifact of the conformal frame chosen.

For sake of completeness it should be mentioned that in addition to the 1-parameter family of solutions, labelled by $M$, isolated solutions may exist, so-called constant dilaton vacua, which have to obey $X=X_{\rm CDV}=\rm const.$ with $w^\prime(X_{CDV}) = 0$. The corresponding geometry has constant curvature, i.e., only Minkowski, Rindler or (A)dS are possible spacetimes for constant dilaton vacua. Incidentally, for the generic case \eqref{eq:EF} the value of the dilaton on an extremal Killing horizon is also subject to these two constraints.

\enlargethispage{1cm}

\subsubsection{A brief review of quantum dilaton gravity with matter}

Adding a matter action for a scalar field $\phi$, coupled to the dilaton via $F(X)$,
\begin{equation}
  \label{eq:vbh:matteraction}
  L^{(m)} = \frac{1}{2} \int_{\mathcal{M}_2} \, F(X) d\phi \wedge 
\ast d\phi =  \int_{\mathcal{M}_2} \extd^2x\sqrt{-g}{\cal{L}}^{(m)}\,,
\end{equation}
to \eqref{eq:FOG} makes the theory non-topological. Hamiltonian analysis yields primary and secondary first class constraints. The latter ($G_i$) form a non-trivial algebra with respect to the Poisson-bracket \cite{Katanaev:2000kc,Grumiller:2001ea}:\footnote{It should be observed that the algebra closes on $\de$, rather than $\de^\prime$; nevertheless, by combining the constraints linearly in a certain way one may obtain an algebra closing with $\de^\prime$, namely the Virasoro algebra (times an abelian one corresponding to Lorentz transformations, cf.~e.g.~\cite{Katanaev:1993fu}).}
\begin{eqnarray}
\hspace{-1truecm} && \left\{ G_1(x^1), G_2({x^1}^\prime) \right\} = - G_2 \,\de(x^1-{x^1}^\prime)\,, \label{algebra1} \\
\hspace{-1truecm} && \left\{ G_1(x^1), G_3({x^1}^\prime) \right\} = G_3 \,\de(x^1-{x^1}^\prime)\,, \label{algebra2} \\
\hspace{-1truecm} && \left\{ G_2(x^1), G_3({x^1}^\prime) \right\} = - \frac{\extd {\mathcal V}}{\extd X^i} G_i \,\de(x^1-{x^1}^\prime)
+\frac{\extd \,\ln{F}}{\extd X}{\cal{L}}^{(m)} G_1 \,\de(x^1-{x^1}^\prime)\,. \label{algebra3}
\end{eqnarray}
Here, $X^i$ denotes $(X,X^+,X^-)$ and $x^1$ is one of the world-sheet coordinates (the one that has not been used as ``time''). 
The simpler case $F=\rm const.$ (minimal coupling) had already been studied before in ref.~\cite{Kummer:1998zs}. A BRST analysis reveals that the BRST charge is nilpotent at ``Yang-Mills level'', i.e., without higher order ghost terms; thus, retrospectively, one may use instead the simpler Faddeev-Popov prescription.
Another crucial observation is that the geometric part of the constraints is {\em linear} in the Cartan variables. Together with a gauge-fixing fermion that implies Eddington-Finkelstein gauge 
\begin{equation}
  \label{eq:EFgf}
  \om_0=0\,,\quad e_0^-=1\,,\quad e_0^+=0\,,
\end{equation}
these features allow an exact path integral quantization, i.e., schematically\footnote{The term ``ghosts'' denotes the whole ghost and gauge-fixing sector. It should be noted that the path integral \eqref{eq:pifull} involves positive and negative values of the dilaton and both orientations ${\rm sign}\, e_1^+=\pm$. Further details on the quantization procedure may be found in appendix E of \cite{Grumiller:2001ea} and in Section 7 of \cite{Grumiller:2002nm}.} 
\begin{multline}
W({\rm sources})=\int \mathcal{D}e_\mu^a  \,\mathcal{D}\omega_\mu \,\mathcal{D}X^i \,\mathcal{D}(\rm ghosts)\,\mathcal{D}\phi\, \\
\times\exp\left[ i\int \extd^2x \left(L_{\rm eff}(e_\mu^a, \omega_\mu, X^i, {\rm ghosts}, \phi) + {\rm sources} \right) \right]\,,
\label{eq:pifull}
\end{multline}
of all fields but matter {\em without} introducing a fixed background geometry. Thus, the quantization procedure is non-perturbative and background independent.
However, there are ambiguities coming from integration constants the fixing of which selects a certain asymptotics of spacetime; two of them are trivial while the third one essentially determines the ADM mass (whenever this notion makes sense).\footnote{The issue of mass is slightly delicate in gravity. For a clarifying discussion in 2D see \cite{Liebl:1997ti}. One of the key ingredients is the existence of the conserved quantity $M$ in \eqref{eq:c} \cite{Banks:1991mk} 
which has a deeper explanation in the context of first order gravity \cite{Grosse:1992vc} and PSMs \cite{Schaller:1994es}. A recent mass definition extending the range of applicability of \cite{Liebl:1997ti} may be found in appendix A of \cite{Grumiller:2004wi}.} Thus, background independence holds only in the bulk but fails to hold in the asymptotic region; we regard this actually as an advantage for describing scattering processes because there is no ``background independent asymptotic observer''.

What one ends up with is a generating functional for Green functions depending solely on the matter field $\phi$, the corresponding source $\sigma$ and on the integration constants mentioned in the previous paragraph \cite{Grumiller:2002dm}:\footnote{$(\mathcal{D}\phi)$ denotes path integration with proper measure. In the context of VBHs questions regarding the measure and 
source terms for geometry are mostly irrelevant. Therefore, 
the generating functional for
Green functions simplifies considerably as compared to the exact case 
\cite{Kummer:1998zs,Grumiller:2002nm}.}
\begin{eqnarray}
&&W(\si ) = \int (\mathcal{D}\phi) \exp{(i L^{\rm eff})}\,,
\label{eq:4.52}\\
&&L^{\rm eff}=\int \extd^2x\left[F(\hat{X})\partial_0\phi\partial_1\phi - \tilde{g}w^\prime(\hat{X}) + \si\phi \right]\,.
\label{eq:4.53a}
\end{eqnarray}
The constant $\tilde{g}$ is an effective coupling which turns out to be inessential and may be absorbed by a redefinition of the unit of length.
For minimal coupling ($F=\rm const.$) $w^\prime$ is the only source for matter vertices. It is a non-polynomial function, in general. Moreover, the quantity $\hat{X}$, which is the quantum version of the dilaton $X$, depends not only on integration constants but also non-locally on matter; to be more precise, it depends non-locally on $(\partial_0\phi)^2$. Thus, in general {\em the effective action \eqref{eq:4.53a} is non-local and non-polynomial in the matter field.} 

\subsubsection{Emergence of VBHs}

A consequence of the quantization procedure sketched above is the possibility to reconstruct geometry from matter. That is, if one had an exact solution to the effective Eqs.~of motion following from \eqref{eq:4.53a}, one obtained not only the behavior of the matter field but simultaneously the geometry on which it propagates by solving relatively simple constraints. In general \eqref{eq:4.52}, \eqref{eq:4.53a} cannot be treated exactly but only perturbatively. Despite of the perturbative treatment {\em no a priori split of geometry into background and fluctuations is invoked}. Rather, to each order in perturbation theory geometry may be reconstructed self-consistently up to the same order, including back reactions. In the following it will be outlined briefly how to obtain contributions from the lowest non-trivial order in matter without going into technical details \cite{Grumiller:2000ah}, i.e., how to obtain the non-local 4-point vertices and the corresponding VBH geometries.\footnote{\label{fn:8} In a perturbative treatment of \eqref{eq:4.52}, \eqref{eq:4.53a} vertices with an arbitrary number of external $\partial_0\phi$ legs are created; in addition, there may be a single $\partial_1\phi$ leg provided $F(\hat{X})\neq \rm const$. Note that the total number of external legs always is even. Thus, to lowest non-trivial order in a perturbative expansion in powers of $\phi$ there are two 4-point vertices, one with four $\partial_0\phi$ legs and one with three $\partial_0\phi$ legs and a $\partial_1\phi$ leg. Non-locality implies their dependence on two sets of coordinates, $x,y$.} 

Of course, one can apply the straightforward but somewhat tedious standard methods to derive their Feynman rules \cite{Kummer:1998zs}. Fortunately, there is an equivalent, albeit much easier, way to derive the Feynman rules, namely by considering matter localized in the following way
\begin{equation}
  \label{eq:matterlocalized}
  (\partial_0\phi)(\partial_1\phi)=c_1\delta^{(2)}(x-y)\,,\quad(\partial_0\phi)^2=c_0\delta^{(2)}(x-y)\,,
\end{equation}
and by solving the {\em classical} Eqs.~of motion up to lowest order in $c_i$. It can be shown \cite{Kummer:1998zs,Grumiller:2002nm} that this mimics the effects of functional differentiation. So in short, instead of taking the $n^{th}$ functional derivative of the generating functional \eqref{eq:4.52} with respect to bilinear combinations of the scalar field -- the brute force method to obtain the Feynman rules for the vertices -- one may localize matter on $n$ points. For lowest order one point is sufficient. Consequently, it turns out that the conserved quantity \eqref{eq:c}, which in the absence of matter determines the BH mass, no longer is constant. In fact, it is not even local due to interactions with matter but rather a function depending on two points $x^i, y^i$ on the world-sheet \cite{Grumiller:2002dm}: 
\begin{equation}
  \label{eq:cvbh}
  M\to M_{\rm eff}(x,y) = M - 2c_0F(y^0)(M+w(y^0))\theta(y^0-x^0)\de(y^1-x^1)\,.
\end{equation}
Thus, even if $M=0$ there may be intermediate states with $M_{\rm eff}\neq 0$. These intermediate states have been called VBHs in ref.~\cite{Grumiller:2000ah}.

Why is it justified to infer the production of VBHs from \eqref{eq:cvbh}? First of all, $M_{\rm eff}$ is an off-shell quantity because clearly the field configuration in Eq.~\eqref{eq:matterlocalized} does not satisfy the Eqs.~of motion. Moreover, this quantity does not influence directly the asymptotics $x^0\to\infty$ because of the $\theta$-function. So the attribute ``virtual'' is adequate. In addition, the classical interpretation of $M$ is as mass of a BH, so $M$ plays a role similar to the charge in electrodynamics (there, when off-shell particles with a certain charge appear they are referred to as virtual ones). 
To settle this issue convincingly one has to reconstruct geometry as outlined above and to check whether or not it corresponds to something resembling a BH. Using a simple coordinate transformation $\extd r\propto I(x^0)\extd x^0$, $\extd u\propto\extd x^1$ the general result is 
\begin{equation}
  \label{eq:ds1}
   \extd s^2_{\rm VBH}=2\extd r\extd u+K(r,u;r_0,u_0)\extd u^2\,,
\end{equation}
with some complicated expression for $K(r,u;r_0,u_0)$ that may be found explicitly in ref.~\cite{Grumiller:2002dm}.
For spherically reduced gravity \eqref{eq:ds1} simplifies to \cite{Fischer:2001vz}
\begin{equation}
\extd s^2_{\rm VBH} = 2 \extd r \extd u + \left(1 - \frac{2M(r,u;r_0,u_0)}{r} - a(r,u;r_0,u_0) r + d(r,u;r_0,u_0)\right) \extd u^2\,.
\label{ds2}
\end{equation}
Remarkably, this looks like the Schwarzschild metric with a Rindler term. Therefore also the notion of ``black hole'' is justified.
The quantities $M$ (essentially given by \eqref{eq:cvbh}), $a$ and $d$ are localized\footnote{The localization of ``mass'' and ``Rindler acceleration'' on a light-like cut is not an artifact of an accidental gauge choice, but has a physical interpretation in terms of the Ricci-scalar \cite{Grumiller:2001rg}. Incidentally, the Ricci-scalar is such that the Einstein-Hilbert action in D=4 vanishes for all VBH configurations.} 
on the cut $u=u_0$ with compact support $r<r_0$. 

The non-local vertices consist of integrals over both sets of coordinates with an integrand containing both pairs of external matter legs at different points $x$ and $y$, and a non-local kernel producing the VBHs (cf.~Fig.~\ref{fig:vertices} below). For instance, the integrated vertex with no $\partial_1\phi$ leg reads
\begin{equation}
  \label{eq:vertex}
  V^{(4)}_{\rm sym} = \int_x\int_y(\partial_0\phi(x))^2\,V^{(4)}_a(x,y)\,(\partial_0\phi(y))^2\,.
\end{equation}
A similar expression holds for the integrated vertex with a $\partial_1\phi$ leg, with the kernel denoted by $V^{(4)}_b$.
The explicit form of the kernels $V^{(4)}_a$, $V^{(4)}_b$ for spherically reduced gravity may be found in ref.~\cite{Fischer:2001vz}, while the general case is derived in \cite{Grumiller:2002dm}.

\begin{figure}
\center
\epsfig{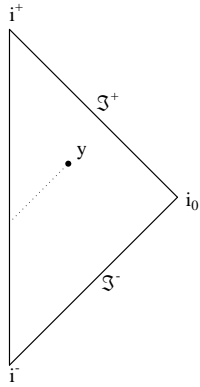}
\caption{CP diagram of a single VBH; the point $y$ corresponds to $u=u_0$, $r=r_0$ in (\ref{ds2})}
\label{fig:1}
\end{figure}
\begin{figure}
\centering
\epsfig{file=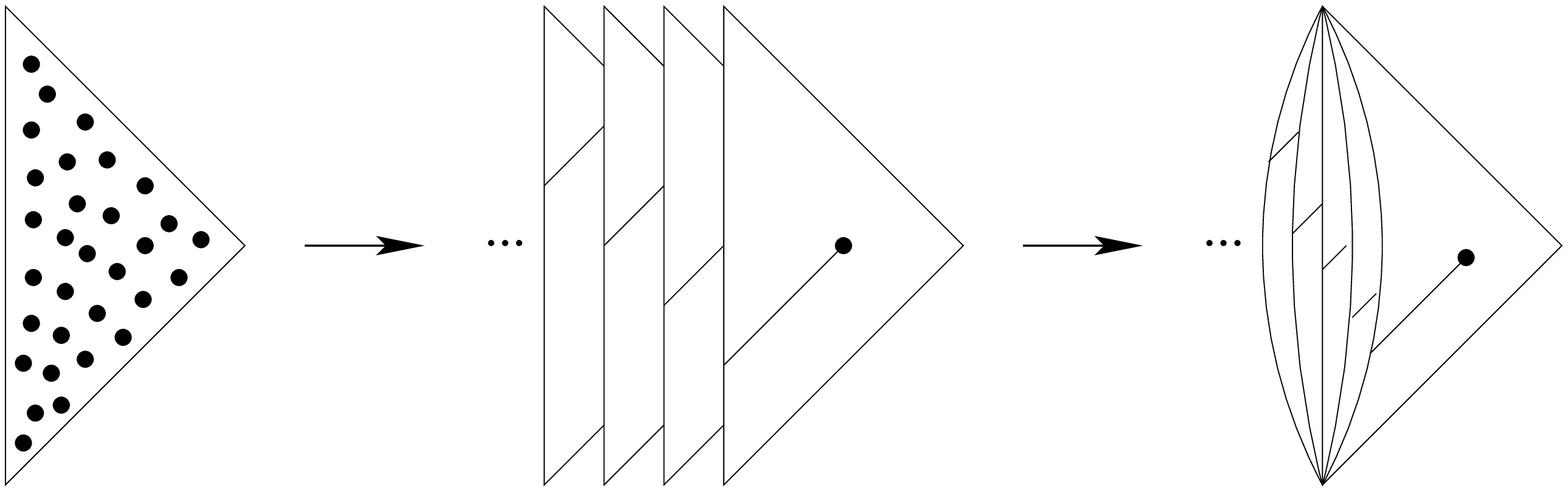,width=0.8\linewidth}
\caption{Constructing the CP diagram of all VBHs}
\label{fig:2}
\end{figure}

A Carter-Penrose (CP) diagram corresponding to the coherent sum of all VBHs can be constructed as follows (cf.\ figs.\ \ref{fig:1}-\ref{fig:2}; all symbols have their standard meaning, i.e., $i^0,i^\pm$ and $\scri^\pm$ are spatial, time-like and light-like infinity, respectively) \cite{Grumiller:2001rg,Grumiller:2003ez}:
\blist
\item Take Minkowski spacetime (or whatever corresponds to the geometry implied by the boundary conditions imposed on the auxiliary fields $X^i$) and draw $N$ different points in its CP diagram; see left diagram of Fig.~\ref{fig:2}.
\item Draw $N$ copies of this CP diagram and add one light like cut to each (always ending at a different point); remove the other $N-1$ points; the line element is given by (\ref{ds2}); see middle diagram of Fig.~\ref{fig:2}. Note: each of these CP diagrams is equivalent to the one depicted in Fig.~\ref{fig:1} with varying endpoint $y$, which is nothing but the CP diagram associated with a {\em single} VBH with line element \eqref{ds2}.
\item Glue together all CP diagrams at $\scri^\pm$ and $i^0$ (which is a common boundary to all these diagrams); see right diagram of Fig.~\ref{fig:2}.
\item Take the limit $N\to\infty$. Thus, the full CP diagram consists of infinitely many layers, each of which resembling Fig.~\ref{fig:1}, the only difference being the end point $y$. Asymptotically all layers coincide. This is a pictorial realization of Everett's ``Many world interpretation''.
\elist 
One should not take the effective geometry at face value -- this would be like 
over-interpreting the role of virtual particles. 
Nonetheless, the simplicity of this geometry and the 
fact that all possible configurations are summed over are nice qualitative features of 
this picture. Because all VBH geometries coincide asymptotically the boundaries of the diagram, $i^0$, $i^\pm$ and $\scri^\pm$, behave in a classical way\footnote{Clearly the boundary conditions imposed play a crucial role in this context. They produce effectively a fixed background, but only at the boundary.}, thus enabling one to construct an ordinary Fock space like in fixed background QFT. Heuristically, the more one zooms into geometry the less classical it becomes.

The situation is complementary to Kucha\v{r}'s proposal of geometrodynamics\footnote{This approach considers only the matterless case and thus a full comparison to our results is not possible.} of BHs: while we have fixed boundary conditions for $X^i$ (and hence a fixed ADM mass) but a ``smeared geometry'' (in the sense that a continuous spectrum of asymptotically equivalent VBHs contributes to the S-matrix), Kucha\v{r} encountered a ``smeared mass'' (obeying a Schr\"odinger Eq.) but an otherwise fixed geometry \cite{Kuchar:1994zk}.

Below it will be shown how the VBHs described above enter the S-matrix together with some consequences which are observable, at least in principle.

\enlargethispage{0.5cm}

\section{VBHs in scattering experiments}\label{se:3}

\subsection{Low scale quantum gravity}\label{se:3.1}

In the past years the possibility of BH production at future colliders \cite{Argyres:1998qn}, 
like LHC \cite{Dimopoulos:2001hw}, and in cosmic rays has been studied in great detail (for reviews cf.~e.g.~\cite{Cavaglia:2002si}). 
A necessary ingredient to experimental verification is the assumption of low-scale quantum gravity, where ``low'' refers to about $1\,TeV$ \cite{Arkani-Hamed:1998rs}. 
If one considers this scenario seriously one should also contemplate the possibility of VBH production. In fact, even if it turned out that the scale explored at LHC is slightly below the quantum gravity scale, and thus real BHs may not be produced with a rate sufficient for detection, in principle effects from VBH production could still be accessible experimentally. As compared to the excitement caused by real BH production the number of studies devoted to VBH production is small. Apart from considerations regarding proton decay\footnote{Since BHs may be responsible for the violation of global quantities such as baryon or lepton number \cite{Zeldovich:1976vq} there was some concern that VBHs might rule out the possibility of $TeV$ range quantum gravity due to proton decay \cite{Adams:2000za} which was refuted in \cite{Kobakhidze:2001yk}.} the only work I am aware of is an unpublished {\tt e-print} \cite{Uehara:2002cj}.\footnote{In ref.~\cite{Uehara:2002cj} a Hawking temperature is assigned to VBHs and effects from Hawking radiation are calculated. This is hard to justify for genuine VBHs but might apply to ``nearly virtual'' BHs.} Thus, regarding VBH production at future colliders or in cosmic rays it seems that there are many issues of potential phenomenological interest awaiting to be discovered.

\subsection{S-matrix for s-wave gravitational scattering}\label{se:3.2}

The idea that BHs must be considered in the S-matrix 
together with 
elementary matter fields has been put forward some time ago 
\cite{'tHooft:1996tq}. The approach \cite{Kummer:1998zs,Grumiller:2000ah,Fischer:2001vz,Grumiller:2001ea,fischervertices,Grumiller:2001rg,Grumiller:2002dm,Grumiller:2003dh,Grumiller:2003sk,Grumiller:2003ez} 
reviewed here, for the first time allowed to derive (rather than to 
conjecture) the appearance of VBH states in the quantum scattering matrix 
of gravity and to predict consequences for certain physical observables.

\begin{figure}
\centering
\epsfig{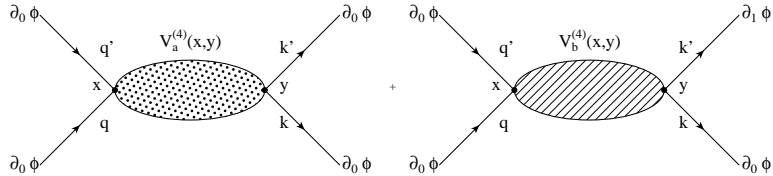}
\caption{The total $V^{(4)}$-vertex (with outer legs) contains a symmetric
contribution $V^{(4)}_a$ and (for non-minimal coupling) a non-symmetric one
$V^{(4)}_b$. The shaded blobs depict the intermediate interactions with VBHs.}
\label{fig:vertices}
\end{figure}

\begin{figure}
\centering
  \epsfig{file=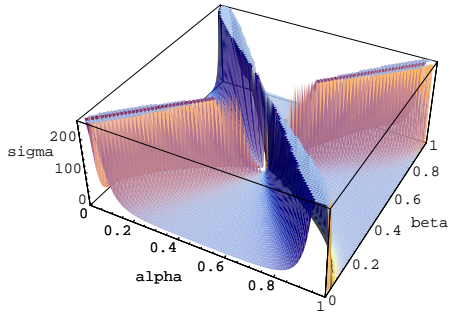,width=.47\linewidth}
  \caption{Kinematic plot of $s$-wave cross-section $d\si/d\al$ for constant $E$}
  \label{fig:kin}
\end{figure}

Qualitatively it is clear what has to be done in order to obtain the S-matrix: 
Take all possible VBHs of Fig.~\ref{fig:2} and sum them coherently with proper
weight factors and suitably attached external legs of scalar fields. To be more precise, one has to take the vertex \eqref{eq:vertex}, calculate the kernel $V^{(4)}_a$ and perform a mode decomposition of the scalar field (which is well-defined because the asymptotic region allows the construction of a standard Fock space), thus introducing creation and annihilation operators $a^\pm_k$ obeying $[a^-_k,a^+_{k^\prime}]\propto\delta(k-k^\prime)$. Then do the same for the non-symmetric vertex and calculate the amplitude for scattering of two ingoing modes with momenta $q,q'$ into two outgoing ones with momenta $k,k'$:
\begin{equation}
  \label{eq:vacamp}
  T(q,q^\prime;k,k^\prime) \propto \langle0|a_k^-a_{k^\prime}^-\left(V^{(4)}_{\rm sym}+V^{(4)}_{\rm non-sym}\right)a_q^+a_{q^\prime}^+|0\rangle
\end{equation}
This had
been done  quantitatively \cite{Fischer:2001vz} in a straightforward but rather lengthy calculation \cite{Grumiller:2001ea,fischervertices}. The physical model behind these detailed calculations is Einstein gravity in D=4, minimally coupled to a massless Klein-Gordon field, truncated to the s-wave sector. Therefore, spherical reduction may be applied and a 2D model of the type discussed above emerges.  
Thus, one is able to study gravitational scattering of matter s-waves in the framework reviewed above. For that model to lowest non-trivial order the tree-graph S-matrix \eqref{eq:vacamp} is given by 
\begin{equation}
T(q, q'; k, k') = -\frac{i\kappa\delta\left(k+k'-q-q'\right)}{2(4\pi)^4 
|kk'qq'|^{3/2}} E^3 \tilde{T}\,, \label{eq:fullamplitude}
\end{equation}
with the total energy $E=q+q'$, $\kappa=8\pi G_N$, 
\begin{multline}
\tilde{T} (q, q'; k, k') := \frac{1}{E^3}{\Bigg [}\,\Pi\, \ln{\frac{\Pi^2}{E^6}}
\\ + \frac{1} {\Pi} \sum_{p \in \left\{k,k',q,q'\right\}}  
p^2 \ln{\frac{p^2}{E^2}} {\Bigg (}3 kk'qq'-\frac{1}{2}
\sum_{r\neq p} \sum_{s \neq r,p}\left(r^2s^2\right){\Bigg )} {\Bigg ]}\,,
\label{eq:amplitude}
\end{multline}
and the momentum transfer function
\begin{equation}
  \label{eq:pi}
  \Pi(k,q,k^\prime) = (k+k^\prime)(k-q)(k^\prime-q)\,.
\end{equation}
Here are some remarks regarding the result \eqref{eq:fullamplitude}-\eqref{eq:pi}:
\paragraph{Scale independence} The interesting part of the scattering amplitude is encoded in the scale independent (!) factor $\tilde{T}$ in \eqref{eq:amplitude}. This issue will be addressed in more detail below. Note that scale invariance does {\em not} apply to the full amplitude \eqref{eq:fullamplitude}.
\paragraph{Forward scattering} The forward scattering poles occurring for $\Pi=0$ should be noted. Their appearance may have been anticipated on general grounds from classical scattering theory.
\paragraph{Simplicity} As a brief glance at the details shows \cite{Grumiller:2001ea,fischervertices} there are actually two contribution to the amplitude which have to be added (cf.~footnote \ref{fn:8} and see Fig.~\ref{fig:vertices}). 
Each of them is not only vastly more complicated than \eqref{eq:amplitude} but also divergent. These somewhat miraculous cancellations urgently ask for some explanation. The one we have found to be convincing is gauge-independence of the S-matrix. Thus, the complicated expressions for single Feynman-diagrams are an artifact of our gauge choice \eqref{eq:EF} which has been a prerequisite for the exact path integration over geometric degrees of freedom, auxiliary fields and ghosts. It remains a challenge to find a simpler derivation of \eqref{eq:amplitude}.
\paragraph{Scattering on self-energy and decay of s-waves} Physically the s-waves of the massless Klein-Gordon field are scattered on their own gravitational self-energy. By rearrangement of the outer legs also a decay of an ingoing s-wave into three outgoing ones is possible and the corresponding decay rate may be calculated \cite{Fischer:2001vz,Grumiller:2001ea,fischervertices}.
\paragraph{CPT invariance} By switching from outgoing to ingoing Eddington-Finkelstein gauge it has been argued in ref.~\cite{Grumiller:2001rg} that the amplitude is CPT-invariant. This is a non-trivial feature because one might expect CPT violation from interactions with VBHs on general grounds and because the effective action \eqref{eq:4.53a} is non-local.
\paragraph{Cross section} With the definitions\\
  \begin{equation}
    \label{eq:momdefs}
    k=E\al\,,\quad k'=E(1-\al)\,\quad q=E\be\,,\quad q'=E(1-\be)\,,
  \end{equation}
where $\al,\be\in[0,1]$ and $E\in\mathbb{R}^+$, a quantity to be interpreted as a 
cross-section for spherical waves can be defined \cite{Fischer:2001vz}:
\begin{equation}
 \frac{d\sigma}{d\alpha}=\frac{1}{4(4\pi)^3}\frac{\kappa^2 E^2 |\tilde{T}
(\alpha, \beta)|^2}{(1-|2\beta-1|)(1-\alpha)(1-\beta)\alpha\beta}\,.
\label{vbh:crosssection}
\end{equation}
The forward scattering poles are clearly visible in Fig.~\ref{fig:kin}.

\enlargethispage{0.5cm}
 
\paragraph{Pseudo self-similarity} Another property discovered and discussed in ref.~\cite{Grumiller:2001rg} is the apparent equivalence of completely different kinematical sectors of the scattering amplitude, i.e., if one zooms into the central region of Fig.~\ref{fig:kin} one obtains another plot which is (almost) identical to that figure. Self-similarity is broken only at next-to-next-to-leading order in an expansion around a reference point close to a forward scattering pole.
\paragraph{VBHs regularize QFT} Often it is claimed that gravity may regularize (some of) the divergences of QFT (for the classical example invoking the self-energy of a charged point particle cf.~e.g.~the first chapter in ref.~\cite{Ashtekar:1991hf}). It is thus of interest that VBHs confirm this hope. To this end consider a generic amplitude in a scattering process in QFT. Typically, in D dimensions one would expect from energy-\-mo\-mentum conservation the appearance of two $\de$-functions in the amplitude, $\de(E-E^\prime)\de(\mbox{\boldmath$\displaystyle\mathbf{p}$}-\mbox{\boldmath$\displaystyle\mathbf{p}$}^\prime)$, where $\mbox{\boldmath$\displaystyle\mathbf{p}$}$ is a D-1 vector, the sum over all ingoing momenta; similarly, $\mbox{\boldmath$\displaystyle\mathbf{p}$}^\prime$ is the sum over all outgoing momenta and $E$ ($E^\prime$) is the sum over all ingoing (outgoing) energies. 
For massless particles in 2D the singular expression $\de(E-E^\prime)\de(0)$ emerges.\footnote{Similar problems arise for the mass zero propagator in 2D \cite{Balasin:1992nm}. Note that the intermediate divergencies mentioned in the paragraph ``Simplicity'' above resemble this type of divergency as they occur for any values of ingoing (outgoing) momenta, but they are ``milder'' than $\de(0)$, namely $\ln{0}$ \cite{Grumiller:2001ea,fischervertices}. So already for individual Feynman diagrams gravity attenuates the divergences, while for the gauge independent amplitude it eliminates them completely.} However, it turns out that for scattering based upon fully quantized gravity \eqref{eq:4.52} one of the $\de$-functions is absent, essentially because of the non-local nature of the interactions mediated by VBHs. Thus, the amplitude \eqref{eq:fullamplitude} contains only one $\de$-function and consequently it is finite.

\paragraph{Symmetry properties of the amplitude} It has been noted already in \cite{Fischer:2001vz,Grumiller:2001ea} that, somewhat surprisingly, the interesting part of the amplitude, $\tilde{T}$, is scale independent, i.e., $\tilde{T}(q,q^\prime;k,k^\prime)=\tilde{T}(\lambda q,\lambda q^\prime;\lambda k,\lambda k^\prime)$ for any non-vanishing $\lambda\in\mathbb{R}$. Here I would like to elaborate on that observation and to discuss also other symmetry properties of the amplitude. To this end it is helpful to introduce the dimensionless quantities $\al$, $\be$ together with the scale $E$ defined in \eqref{eq:momdefs}. In terms of these, together with the abbreviations $A=\al(1-\al)$, $B=\be(1-\be)$, one gets
$\Pi(\al,\be,E)=E^3 \left(A-B\right)$.
Plugging this into \eqref{eq:amplitude} it is simple to show its independence from the scale $E$ by collecting all terms containing $E$: $3\ln{E^2}[A-B+(2AB(1-A-B)-\frac12(2A(1-2B)+2B(1-2A)))/(A-B)]=0$.
Thus, it is established that the interesting part of the scattering amplitude, $\tilde{T}$ as defined in \eqref{eq:amplitude}, does not depend on $E$ at all but only on the kinematic factors $\al$ and $\be$. 
The discussion in ref.~\cite{Moretti:2002mp} suggests to take this scale independence seriously and to look for further symmetries.\footnote{I am grateful to N.~Pinamonti and D.~Vassilevich for discussions on this subject after a talk of the former at the MPI Leipzig. It should be pointed out that the $SL(2,\mathbb{R})$ symmetry regarding the energy spectrum found in ref.~\cite{Moretti:2002mp} relies on scale reparameterization of the total scattering amplitude, which does not apply to \eqref{eq:fullamplitude}.} 
Boosts of all modes $E\to\gamma E$, $p\to\gamma p$ (where $p=k,q,k^\prime,q^\prime,$) just amount to an energy rescaling because particles are massless so the Lorentz angle does not change; therefore, the analysis above applies and boosts are a symmetry of the amplitude. There seem to be no further continuous symmetries.
However, there are discrete symmetries: one may permute within the set of ingoing and/or outgoing particles. Consistently, \eqref{eq:fullamplitude} is invariant under $\al\leftrightarrow(1-\al)$ and $\be\leftrightarrow(1-\be)$. This residual invariance may be fixed trivially by restricting $\al\in[0,1/2]$, $\be\in[0,1/2]$. Exchanging in- with outgoing, i.e., $\al\leftrightarrow\be$ swaps the sign of $\tilde{T}$ and therefore the full amplitude \eqref{eq:fullamplitude} gets complex conjugated, as expected for a theory exhibiting CPT invariance.

\section{On the specific heat of BHs} \label{se:4}

From a thermodynamical point of view the specific heat of the father of all BHs, the Schwarzschild BH, is quite remarkable, namely negative. This leads one to ask the question how quantum corrections due to VBHs influence the specific heat. To this end it is recalled briefly how to obtain the specific heat from entropy for generic dilaton gravity (for a brief review on the latter see above).
The final Subsection is devoted to VBH corrections of the (inverse) specific heat for the CGHS model.

\subsection{BH thermodynamics in 2D dilaton gravity}

Before being able to appreciate the relevance of quantum corrections to the specific heat it is worthwhile to collect a few classical results first. 

\paragraph{Hawking temperature} There are many ways to calculate the Hawking temperature, some of them involving the coupling to matter fields, some of them being purely geometrical. Because of its simplicity we will restrict ourselves to a calculation of the geometric Hawking temperature as derived from surface gravity (cf.~e.g.~\cite{Wald:1999vt}).\footnote{If defined in this way Hawking temperature turns out to be independent of the conformal frame. Although for the main application below only asymptotically flat spacetimes are encountered it should be noted that identifying Hawking temperature with surface gravity is somewhat naive for spacetimes which are not asymptotically flat. But the difference is just a redshift factor and for quantities like entropy or specific heat actually \eqref{eq:ht} is the relevant quantity as it coincides with the period of Euklidean time (cf.~e.g.~\cite{Gibbons:1994cg}).  \label{fn:hp}} 
The latter can be calculated by taking the normal derivative $\extd/\extd X$ of the Killing norm $K(X;M)$ evaluated on one of the Killing horizons $X=X_h$, where $X_h$ is a solution of $K(X_h;M)=0=(M+w(X_h))$, thus yielding
\begin{equation}
  \label{eq:ht}
  T_H =  \frac{1}{2\pi} \Big|w'(X) \Big|_{X=X_h}\,.  
\end{equation}
The numerical prefactor in \eqref{eq:ht} can be changed e.g.~by a redefinition of the Boltzmann constant. It has been chosen in accordance with refs.~\cite{Kummer:1999zy,Grumiller:2002nm}.

\paragraph{Entropy} 
In 2D dilaton gravity there are various ways to calculate the Beken\-stein-Hawking entropy \cite{Bekenstein:1973ur}. 
Using two different methods Gegenberg, Kunstatter and Louis-\-Martinez were able to calculate the entropy for rather generic 2D dilaton gravity \cite{Gegenberg:1995pv}. Later, Cadoni and Mignemi confirmed their result for a particular model using the Cardy formula and counting microstates\footnote{As opposed to String Theory, where the microstates are $D$-branes (for reviews cf.~\cite{Youm:1997hw}), 
or to Loop Quantum Gravity, where the microstates are quanta of area (for a review cf.~e.g.~\cite{Ashtekar:2000eq}), it is fair to say that in the context of 2D dilaton gravity it is not quite clear what these microstates actually are -- cf.~e.g.~the recent discussion in ref.~\cite{Carlip:2004mn}. If one employs Hod's conjecture \cite{Hod:1998vk} one is led to consider quasi-normal modes in the limit of high damping. This has been performed recently by Kettner, Kunstatter and Medved \cite{Kettner:2004aw} (cf.~Eq.~(36) in that ref.). Their result is remarkable insofar as it is rather insensitive to geometry and depends solely on the scale set by surface gravity and on the way matter is coupled to the dilaton field. Thus, taking Hod's conjecture seriously, it appears that one cannot avoid to conclude that the microstates in 2D dilaton gravity are built from matter degrees of freedom. This is in accordance with the theory being topological in the absence of matter, but it does not explain why the derivations of entropy which do not employ matter at all work so well. So alternatively, one might conclude that Hod's conjecture is not applicable (to 2D dilaton gravity). I thank G.~Kunstatter for correspondence on \cite{Kettner:2004aw}.} \cite{Cadoni:1998sg}. 
Finally, Carlip rederived the general result by applying CFT methods \cite{Carlip:1999cy} and later also by virtue of the Cardy formula \cite{Carlip:2002be}. These considerations are based upon earlier observations regarding near horizon conformal symmetry for the Schwarzschild BH \cite{Carlip:1998wz}. 
Surprisingly enough, the naive derivation which employs only the thermodynamic relation $\extd S = \extd M / T$ yields the correct result: Entropy equals the dilaton field evaluated at the Killing horizon,\footnote{Up to a multiplicative constant which may be absorbed by a redefinition of Newton's constant.}
\begin{equation}
  \label{eq:entropy}
  S = 2\pi X_h\,.
\end{equation}

\paragraph{Specific heat}
By virtue of $C_s=T\extd S/\extd T$ the inverse specific heat reads
\begin{equation}
  \label{eq:inversecs}
  C_s^{-1} = \frac{1}{2\pi} \left. \frac{\extd}{\extd X}\ln{w'(X)}\right|_{X=X_h}\,.
\end{equation}
It is also independent of the conformal frame, but in the simple frame $I=\rm const.$ an intriguing reformulation exists:
\begin{equation}
  \label{eq:csother}
  C_s = \frac{8\pi^2s}{-r_h} T_H\,,
\end{equation}
where $r_h$ is given by \eqref{eq:curv} evaluated on the horizon. The sign $s=\pm 1$ is positive if $w'$ is negative on the horizon, otherwise it is positive.\footnote{If $w'$ vanishes on the horizon then $T_H=0$ and $s$ is irrelevant, unless simultaneously $w''=0$. In that special case -- which arises if and only if the Killing norm has at least a triple zero -- it is better to use \eqref{eq:inversecs}.} In all examples below $s=+1$. Thus, the sign of $r_h$ defines whether the specific heat is positive (e.g.~for AdS) or negative (e.g.~for dS). On a curious sidenote it is mentioned that \eqref{eq:csother} behaves like an electron gas at low temperature with Sommerfeld constant $\gamma=8\pi^2s/(-r_h)$. However, this analogy does not go too far because the (Planck version of the) third law of thermodynamics is not fulfilled necessarily, i.e., entropy need not vanish as $T_H\to 0$ -- for instance, the only model of the whole $ab$-family in Fig.~\ref{tab:ex} which obeys the third law, and consequently $S=C_s$, is the Jackiw-Teitelboim model.

\begin{figure}
\hspace{-1.4truecm}
{\footnotesize
\begin{tabular}{|l||>{$}c<{$}|>{$}c<{$}|>{$}c<{$}|>{$}c<{$}|>{$}c<{$}|} 
\hline
Model & w(X) & X_h=S/(2\pi) & T_H & C_s \\ \hline \hline
Schwarzschild BH \cite{Schwarzschild:1916uq} & -\lambda\sqrt{X} & M^2/\lambda^2 & \lambda^2/(4\pi M)& -\lambda^2/(4\pi T_H^2) \\
Witten BH/CGHS \cite{Mandal:1991tz}
& -\lambda X  & M/\lambda & \lambda/(2\pi) & \infty \\
Jackiw-Teitelboim \cite{Teitelboim:1983ux}
& -\lambda X^2 & \sqrt{M/\lambda} & \sqrt{\lambda M}/\pi & 2\pi^2 T_H/\lambda \\ 
$ab$-family \cite{Katanaev:1997ni} & -\lambda X^{b+1} &  (M/\lambda)^{1/(b+1)} & \alpha(M/\lambda)^{b/(b+1)} & 2\pi b^{-1} (T_H/\alpha)^{1/b} \\
Schwarzschild-AdS \cite{Hawking:1982dh} & -\lambda\sqrt{X}(1+X/\ell^2) & \rm soluble\,\,alg. & \frac{\lambda}{4\pi}\sqrt{1/X_h}(1+3X_h/\ell^2) & -4\pi X_h\frac{1+3X_h/\ell^2}{1-3X_h/\ell^2}\\
Reduced CS \cite{Guralnik:2003we}
& -\lambda(X^2-c)^2 & \sqrt{c+\sqrt{M/\lambda}} & \frac{2\lambda}{\pi}\sqrt{M/\lambda}\sqrt{c+\sqrt{M/\lambda}} & \frac{\pi^2}{\lambda}T_H/(3X_h^2-c) \\ 
\hline
\end{tabular}
}
\caption{Table of examples. Note the (irrelevant) scale factor $\lambda>0$ and the abbreviation $\alpha=\lambda(b+1)/(2\pi)$. For simplicity $X$ is assumed to be positive. In the penultimate example $X_h$ is given uniquely by the real root of a cubic Eq. In the last example all expressions refer to the outermost horizon. In the first five examples horizons exist iff $M>0$, in the last one iff $M\geq 0$.}
\label{tab:ex}
\end{figure} 

\paragraph{Free energy et al.} Once entropy $S$ is known as a function of temperature $T$ and energy $M$ due to the absence of pressure it is straightforward to calculate other thermodynamical quantities of interest. For instance, the free energy is given by $F=M-TS$; the Euklidean action follows from $I=F/T$; the partition function is given by $Z=e^{-I}$. If more than one horizon is present one can assign an entropy to each of them, but of course the thermodynamical discussion of the whole system becomes more complicated.

\paragraph{Hawking-Page like phase transition} In their by now classic paper on thermodynamics of BHs in AdS, Hawking and Page found a critical temperature signalling a phase transition between a BH phase and a pure AdS phase \cite{Hawking:1982dh}. This has engendered much further research, mostly in the framework of the AdS/CFT correspondence (for a review cf.~\cite{Aharony:1999ti}). This transition is displayed most clearly by a change of the specific heat from positive to negative sign: for Schwarzschild-AdS according to Fig.~\ref{tab:ex} the critical value of $X_h$ is given by $X_h^c=\ell^2/3$. For $X_h>X_h^c$ the specific heat is positive, for $X_h<X_h^c$ it is negative.\footnote{Actually, in the original work \cite{Hawking:1982dh} Hawking and Page did not invoke the specific heat directly. The consideration of the specific heat as an indicator for a phase transition is in accordance with the discussion in ref.~\cite{Brown:1994gs}.} By analogy, a similar phase transition may be expected for other models with corresponding behavior of $C_s$. For instance, the last example in Fig.~\ref{tab:ex} exhibits also a critical value of $X_h$, namely $X_h^c=\pm\sqrt{c/3}$. Notably, this value can never be reached for outer horizons, but may be reached for inner ones. 

By virtue of the reformulation \eqref{eq:csother} a candidate for a Hawking-Page like transition arises at a certain critical value of $M\in(0,\infty)$ such that $r_h=0$ at the (non-extremal) Killing horizon.

\paragraph{Summary} The function $w(x)+M$ encodes all thermodynamical properties discussed above: its zeros yield the value of the dilaton at horizons and thus entropy, its first derivative is proportional to the Hawking temperature, and its second derivative (together with the first) determines the specific heat. If $w''$ vanishes on the horizon for a finite value of $M$ a Hawking-Page like phase transition may occur.

\subsection{Quantum corrections to the specific heat}

In general, quantum corrections are, well, corrections to some classical result in the sense that the dominant contribution is classical. However, there exist instances where quantum corrections become pivotal and compete with (or even beat the) classical contribution. 

The table in Fig.~\ref{tab:ex} reveals that for the CGHS model the inverse specific heat vanishes classically. Thus, any non-vanishing contribution to $C_s^{-1}$ must be purely of quantum origin. Therefore, it is of some interest to study these corrections in more detail. This has been undertaken\footnote{Several years before our derivation Zaslavskii has performed a comparable calculation \cite{Zaslavsky:1996dg}. Although the specific heat is not calculated explicitly in that work it is a trivial excercise to extract it from the quantum corrected expressions for mass and temperature. Comparing his result to ours agreement is found up to an overall sign. Extensive discussions and cross-checkings of pesky signs have not revealed any obvious sign error in either of the publications. It should be mentioned, however, that there is actually a difference between our calculation and Zaslavskii's concerning the boundary conditions imposed: while he used Hartle-Hawking boundary conditions we have employed Unruh boundary conditions. It is not clear whether this difference is responsible for the relative sign. In any case, the important conclusion remains unchallenged by such details: interactions with VBHs produce crucial corrections to the inverse specific heat of the CGHS BH.\label{fn:zas}} in ref.~\cite{Grumiller:2003mc} by coupling geometry to a single massless scalar field and peforming path integration over geometry non-perturbatively as outlined in Section \ref{se:2.2}. It has been found that the interaction with matter induced VBHs\footnote{Classically, for the CGHS model VBHs have no observable effect \cite{Grumiller:2002dm}. Note that the VBH interpretation need not be adopted -- indeed, neither \cite{Zaslavsky:1996dg} nor \cite{Grumiller:2003mc} mention this notion explicitly -- but it is in the spirit of the present work.} effectively amounts to a shift of the Killing norm from its classical value $K_{c} = 1-(M/\lambda) e^{-2\lambda r}$ (with $X=\exp{(2\lambda r)}$) to
\begin{equation}
  \label{eq:cghskq}
   K_{q} = 1-\frac{M}{\lambda}e^{-2\lambda r} + \frac{M}{48\pi\lambda}e^{-4\lambda r}\,.
\end{equation}
This implies an effective shift of $w$ from its classical value $w_c=-\lambda X$ to
\begin{equation}
  \label{eq:wcghs}
  w(X)=w_c\left(1+\frac{M}{48\pi\lambda X^2}\right)\,,
\end{equation}
leaving $I=I_c=1/(2\lambda X)$ uncorrected. The results above are valid provided $M\gg\lambda$.
Here are some consequences of (and remarks to) the result \eqref{eq:cghskq}:

\paragraph{Positive specific heat} The most dramatic implication of quantum corrections arises for the inverse specific heat: while it vanishes according to the standard analysis (see Fig.~\ref{tab:ex}), it turns out to be positive when interactions with VBHs are taken into account,
  \begin{equation}
    \label{eq:qcs}
    C_s=\frac{96\pi^2}{\lambda^2} M^2\,.
  \end{equation}
This implies that quantum effects tend to stabilize the system (see, however, footnote \ref{fn:zas}). 

\paragraph{Violation of area law} The behavior of the Killing norm \eqref{eq:cghskq} implies that the horizon is shifted to slightly smaller values of $r$ due to quantum corrections from VBHs. Therefore, Hawking's area theorem \cite{Hawking:1971tu} 
is violated. 

\paragraph{Corrections to radiation loss} Applying the 2D Stefan-Boltzmann law yields to leading order a decrease of the BH mass linear in ``time'', proportional to $T_H^2$. This is modified according to
  \begin{equation}
    \label{eq:massloss}
M(t)\approx M_0-\frac{\pi}{6} (T_H^0)^2(t-t_0)+\frac{\la}{24\pi}\ln{\frac{M(t)}{M_0}} \,,
  \end{equation}
where $t>t_0$, $M_0=M(0)$ and $T_H^0=\lambda/(2\pi)$ (in accordance with Fig.~\ref{tab:ex}). Terms of higher order in $\lambda/M$ have been neglected. The last term in \eqref{eq:massloss} is the one which is due to corrections from VBHs. Note that $M(t)$ may be expressed in terms of the Lambert W-function \cite{Corless:1996}.

\paragraph{Conformal non-invariance} The quantum corrections crucially depend on the conformal frame -- for instance, if one uses the simple conformal frame $I=\rm const.$ then no quantum corrections to the Killing norm arise. The feature of classical conformal invariance of certain quantities but quantum non-invariance is in accordance with the general discussion in \cite{Flanagan:2004bz}.
Fortunately, for the CGHS model there exists a preferred way to choose a conformal frame, a so-called Minkowski ground state frame;\footnote{The definition is as follows: for vanishing value of the mass $M$ geometry must be Minkowski space. This leads to $I\propto X^{-a}$ with $a=b+1$ for the $ab$-family in Fig.~\ref{tab:ex}. Hence, for the CGHS model $a=1$.} it is this frame that has been used in the derivation of \eqref{eq:cghskq}.\footnote{Note that \eqref{eq:cghskq} would allow to consider the following correction as natural $I(X)=I_c(X)(1-1/(48\pi X))$. However, this ``quantum correction of the conformal frame'' is problematic because it induces a singularity in the conformal factor at $48\pi X=1$.}

\paragraph{Logarithmic entropy corrections} By a simple thermodynamical calculation based upon \eqref{eq:cghskq} corrections to the entropy have been calculated in ref.~\cite{Grumiller:2003sk}:
  \begin{equation}
    \label{eq:Scorr}
    S=S_0-\frac{1}{24}\ln{S_0} + \mathcal{O} (1)\,,
  \end{equation}
where $S_0=2\pi M/\lambda$ (cf.~Fig.~\ref{tab:ex}). The logarithmic behavior is in qualitative agreement with the one found in the literature by various methods \cite{Mann:1998hm} 
 (cf.~\cite{Ghosh:2004rq} for a brief and recent summary); the factor $1/24$ is in accordance with \cite{Fiola:1994ir}. 

\section{Outlook}\label{se:5}

Several interesting physical consequences from interactions with VBHs can be deduced, both in the Euklidean and in the Minkowskian approach, as reviewed in Sections 2-4. I conclude briefly with a couple of open issues.

\paragraph{Experimental challenges}

Having established that real BHs are part of Nature it seems natural to consider experiments sensitive to virtual ones. Although the hope has been expressed in Section \ref{se:3.1} that such experiments may be feasible if the quantum gravity scale is at low energies, a thorough study of VBHs in that framework still is lacking. By the same token that real BHs may lead to observable effects at near future colliders or in high energy cosmic rays one could argue that VBHs will imply observable consequences. This might play a pivotal role if the quantum gravity scale turns out to be low, but not low enough to yield convincing evidence for real BHs.

Regarding a verification of the scattering amplitude \eqref{eq:fullamplitude} prospects do not look too promising: one would need a system where all forces but gravity can be neglected, which is spherically symmetric and which consists solely of massless scalar particles. Only the last condition may be dropped with ease in the framework presented here, while the first two are crucial ingredients. The only system which comes to my mind exhibiting similar features consists of rapidly expanding or collapsing spherical shells (not necessarily thin ones), like the s-wave part of a supernova. So in conclusion, it seems difficult to invent an experimental setup which manages to unravel the interesting kinematical features hidden in \eqref{eq:amplitude}, besides the forward scattering poles, but undoubtedly it is a very interesting challenge. 

\paragraph{Theoretical challenges}

One of the reasons why VBHs are so interesting from a theoretical point of view is the puzzle of information loss which arises for real BHs. While Hawking argued some time ago in favor of VBH-induced information loss, studies in 2D revealed that no such information loss occurs. Of course, this does not solve the information paradox for macroscopic BHs; but it shows that microscopic (virtual) BHs enter the S-matrix just like any other particle, a conjecture put forward some time ago by 'tHooft \cite{'tHooft:1996tq}. Nevertheless, there is a link between these microscopic studies and macroscopic considerations like in refs.~\cite{Giddings:2001pt}: in both cases non-locality plays a crucial role. While for (microscopic) VBHs non-locality led to a finite result for the S-matrix \eqref{eq:fullamplitude}, for macroscopic evaporating BHs the violation of the ``locality bound'' \cite{Giddings:2001pt} contradicts the assumption of independent Hilbert spaces for the interior and the exterior of a BH and thus information need not be lost.

Several generalizations of the results are obvious: for instance, it would be interesting to study quantum corrections to the specific heat generically, in particular for the Schwarzschild BH. Extensions to SUGRA exhibit also the VBH phenomenon \cite{Bergamin:2004us}, but so far no amplitudes or corrections to specific heat have been calculated. 
Certainly it would be gratifying to extend the 2D study of Minkowskian VBHs and their effects on the S-matrix to more complicated systems in D=4, i.e., to drop the assumption of spherical symmetry. 

Finally, it is fair to ask whether VBHs exist beyond the Euklidean and the Minkowskian path integral approach addressed in this work.  
In the context of Loop Quantum Gravity (LQG) there has been recent progress in describing quantum horizons for spherically symmetric configurations \cite{Bojowald:2004}. Also, it has been found in that ref.~that a binary degree of freedom exists, essentially the orientation of the spherically symmetric isolated horizon. 
This is a promising step towards VBHs as described in Section \ref{se:2} because also in the path integral \eqref{eq:pifull} we have summed over positive orientations ($e_1^+>0$) and negative ones ($e_1^+<0$). However, the boundary conditions imposed in the asymptotic region uniquely select one of these orientations for the effective line element \eqref{ds2}. Thus, although in intermediate states both orientations are possible, for the fiducial observer in the asymptotic region one orientation is selected. It would be interesting to observe similar features in LQG. Thus, as a next step one could consider an isolated quantum horizon {\em together} with an asymptotic region which should be chosen to be essentially the same for all spin network configurations, e.g.~a flat one or dS space. Then, one would have an asymptotic region behaving almost classically, while the bulk part of geometry, in particular the horizon, still behaved in a quantum way. In this manner, if it turned out to be possible to relax the condition on spherical symmetry, even graviton-graviton scattering with intermediate VBHs might be described by LQG.

For String Theory the answer is affirmative: as the Witten BH/CGHS model follows from strings in D=2 and VBHs arise for that model they may be expected to be a generic feature of String Theory. A concrete realization of VBHs in a more general framework of String Theory, i.e., not restricted to the Witten BH, could be an interesting subject for the future. 
 
\section*{Acknowledgements}

This work has been supported by an Erwin-Schr\"odinger fellowship granted by the Austrian Science Foundation (FWF), project J-2330-N08. I am grateful to my long-time collaborators on 2D gravity, W.~Kummer and D.~Vassilevich, for numerous stimulating discussions and to P.~Fischer for joining our efforts and for providing valuable input. I thank S.~Giddings, G.~Landsberg and D.J.~Schwarz for helpful correspondence and M.~Bojowald for sending me a draft of ref.~\cite{Bojowald:2004} prior to submission to the {\tt arXiv}. Last but not least I render special thanks to D.~Ahluwalia-Khalilova for the kind invitation to compile this review.  
       

\begin{thebibliography}{10%
0}

\bibitem{Casimir:1948dh}
H.~B.~G. Casimir, ``On the attraction between two perfectly conducting
  plates,'' {\em Kon. Ned. Akad. Wetensch. Proc.} {\bf 51} (1948)
793--795.

\bibitem{Bordag:2001qi}
M.~Bordag, U.~Mohideen, and V.~M. Mostepanenko, ``{New developments in the
  Casimir effect},'' {\em Phys. Rept.} {\bf 353} (2001) 1--205,
\href{http://www.arXiv.org/abs/quant-ph/0106045}{{\tt quant-ph/0106045}}.

\bibitem{Fadin:1988fn}
V.~S. Fadin and V.~A. Khoze, ``Production of a pair of heavy quarks in e+ e-
  annihilation in the threshold region,'' {\em Sov. J. Nucl. Phys.} {\bf 48}
  (1988)
309--313;
M.~J. Strassler and M.~E. Peskin, ``{The Heavy top quark threshold: QCD and the
  Higgs},'' {\em Phys. Rev.} {\bf D43} (1991)
1500--1514;
W.~Kummer and W.~M{\"o}dritsch, ``Relativistic bound state equation for
  unstable fermions and the toponium width,'' {\em Phys. Lett.} {\bf B349}
  (1995) 525--532,
\href{http://www.arXiv.org/abs/hep-ph/9501406}{{\tt hep-ph/9501406}}.

\bibitem{Hagiwara:2002fs}
{\bf Particle Data Group} Collaboration, K.~Hagiwara {\em et al.}, ``Review of
  particle physics,'' {\em Phys. Rev.} {\bf D66} (2002)
010001;
{\bf D0} Collaboration, V.~M. Abazov {\em et al.}, ``A precision measurement of
  the mass of the top quark,'' {\em Nature} {\bf 429} (2004) 638--642,
\href{http://www.arXiv.org/abs/hep-ex/0406031}{{\tt hep-ex/0406031}}.

\bibitem{Schuler:1996fc}
G.~A. Schuler and T.~Sjostrand, ``Parton distributions of the virtual photon,''
  {\em Phys. Lett.} {\bf B376} (1996) 193--200,
\href{http://www.arXiv.org/abs/hep-ph/9601282}{{\tt hep-ph/9601282}}.

\bibitem{Schodel:2002py}
R.~Schodel {\em et al.}, ``{A Star in a 15.2 year orbit around the supermassive
  black hole at the center of the Milky Way},'' {\em Nature} {\bf 419} (2002)
694--696.

\bibitem{McClintock:2003gx}
J.~E. McClintock and R.~A. Remillard, ``Black hole binaries,''
\href{http://www.arXiv.org/abs/astro-ph/0306213}{{\tt astro-ph/0306213}}.

\bibitem{Carlip:2001wq}
S.~Carlip, ``{Quantum gravity: A progress report},'' {\em Rept. Prog. Phys.}
  {\bf 64} (2001) 885,
\href{http://www.arXiv.org/abs/arXiv:gr-qc/0108040}{{\tt arXiv:gr-qc/0108040}};
L.~Smolin, ``How far are we from the quantum theory of gravity?,''
\href{http://www.arXiv.org/abs/hep-th/0303185}{{\tt hep-th/0303185}};
E.~Alvarez, ``Quantum gravity,''
\href{http://www.arXiv.org/abs/gr-qc/0405107}{{\tt gr-qc/0405107}}.

\bibitem{Preskill:1992tc}
J.~Preskill, ``Do black holes destroy information?,''
\href{http://www.arXiv.org/abs/hep-th/9209058}{{\tt hep-th/9209058}};
D.~N. Page, ``Black hole information,''
\href{http://www.arXiv.org/abs/hep-th/9305040}{{\tt hep-th/9305040}};
T.~Banks, ``Lectures on black holes and information loss,'' {\em Nucl. Phys.
  Proc. Suppl.} {\bf 41} (1995) 21--65,
\href{http://arXiv.org/abs/hep-th/9412131}{{\tt hep-th/9412131}};
G.~'t~Hooft, ``{Black holes, Hawking radiation, and the information paradox},''
  {\em Nucl. Phys. Proc. Suppl.} {\bf 43} (1995)
1--11;
V.~Frolov and I.~Novikov, {\em {Black Hole Physics}}.
\newblock Kluwer Academic Publishers, 1998.

\bibitem{GR17}
As there is no {\tt e-print} at the time of writing this article cf.~the summary on J.~Baez' webpage \href{http://math.ucr.edu/home/baez/week207.html}{{\tt http://math.ucr.edu/home/baez/week207.html}}.

\bibitem{Hawking:1996ag}
S.~W. Hawking, ``Virtual black holes,'' {\em Phys. Rev.} {\bf D53} (1996)
  3099--3107,
\href{http://www.arXiv.org/abs/hep-th/9510029}{{\tt hep-th/9510029}}.

\bibitem{Grumiller:2000ah}
D.~Grumiller, W.~Kummer, and D.~V. Vassilevich, ``The virtual black hole in 2d
  quantum gravity,'' {\em Nucl. Phys.} {\bf B580} (2000) 438--456,
\href{http://www.arXiv.org/abs/gr-qc/0001038}{{\tt gr-qc/0001038}}.

\bibitem{Grumiller:2002nm}
D.~Grumiller, W.~Kummer, and D.~V. Vassilevich, ``Dilaton gravity in two
  dimensions,'' {\em Phys. Rept.} {\bf 369} (2002) 327--429,
\href{http://arXiv.org/abs/hep-th/0204253}{{\tt hep-th/0204253}}.

\bibitem{wheelerrelativity}
J.~Wheeler, ``Geometrodynamics and the issue of the final state,'' in {\em
  Relativity, Groups and Topology}, C.~DeWitt and B.~DeWitt, eds., p.~316.
\newblock Gordon and Breach, 1964.

\bibitem{Baez:1999sr}
J.~C. Baez, ``{An introduction to spin foam models of BF theory and quantum
  gravity},'' {\em Lect. Notes Phys.} {\bf 543} (2000) 25--94,
\href{http://www.arXiv.org/abs/gr-qc/9905087}{{\tt gr-qc/9905087}};
C.~Rovelli, ``Loop quantum gravity,'' {\em Living Rev. Rel.} {\bf 1} (1998) 1,
\href{http://www.arXiv.org/abs/gr-qc/9710008}{{\tt gr-qc/9710008}}.

\bibitem{Hawking:1988ae}
S.~W. Hawking, ``Wormholes in space-time,'' {\em Phys. Rev.} {\bf D37} (1988)
904--910;
S.~R. Coleman, ``Why there is nothing rather than something: {A} theory of the
  cosmological constant,'' {\em Nucl. Phys.} {\bf B310} (1988)
643.

\bibitem{Hawking:1978zw}
S.~W. Hawking, ``Space-time foam,'' {\em Nucl. Phys.} {\bf B144} (1978)
349--362.

\bibitem{nakaharageometry}
M.~Nakahara, {\em {Geometry, Topology and Physics}}.
\newblock IOP Publishing, Bristol, 1990.

\bibitem{Wall:1964}
C.~T.~C. Wall, ``On simply-connected 4-manifolds,'' {\em J. London Math. Soc.}
  {\bf 39} (1964) 141--149.

\bibitem{Ernst:1976}
F.~J. Ernst, ``Removal of the nodal singularity of the {$C$}-metric,'' {\em J.
  Math. Phys.} {\bf 17} (1976) 515.

\bibitem{Hawking:1997ia}
S.~W. Hawking and S.~F. Ross, ``Loss of quantum coherence through scattering
  off virtual black holes,'' {\em Phys. Rev.} {\bf D56} (1997) 6403--6415,
\href{http://www.arXiv.org/abs/hep-th/9705147}{{\tt hep-th/9705147}}.

\bibitem{Prestidge:1998bk}
T.~Prestidge, ``Higher spin field equations in a virtual black hole metric,''
  {\em Phys. Rev.} {\bf D58} (1998) 124022,
\href{http://www.arXiv.org/abs/hep-th/9802028}{{\tt hep-th/9802028}}.

\bibitem{Garay:1999cy}
L.~J. Garay, ``Quantum evolution in spacetime foam,'' {\em Int. J. Mod. Phys.}
  {\bf A14} (1999) 4079--4120,
\href{http://www.arXiv.org/abs/gr-qc/9911002}{{\tt gr-qc/9911002}}.

\bibitem{Benatti:2001fa}
F.~Benatti and R.~Floreanini, ``Massless neutrino oscillations,'' {\em Phys.
  Rev.} {\bf D64} (2001) 085015,
\href{http://www.arXiv.org/abs/hep-ph/0105303}{{\tt hep-ph/0105303}}.

\bibitem{Benatti:1998vu}
F.~Benatti and R.~Floreanini, ``Non-standard neutral kaons dynamics from
  infinite statistics,'' {\em Annals Phys.} {\bf 273} (1999) 58--71,
\href{http://www.arXiv.org/abs/hep-th/9811196}{{\tt hep-th/9811196}}.

\bibitem{Ellis:1984jz}
J.~R. Ellis, J.~S. Hagelin, D.~V. Nanopoulos, and M.~Srednicki, ``Search for
  violations of quantum mechanics,'' {\em Nucl. Phys.} {\bf B241} (1984)
381;
P.~Huet and M.~E. Peskin, ``Violation of {CPT} and quantum mechanics in the
  {K}0 - anti-{K}0 system,'' {\em Nucl. Phys.} {\bf B434} (1995) 3--38,
\href{http://www.arXiv.org/abs/hep-ph/9403257}{{\tt hep-ph/9403257}};
J.~R. Ellis, J.~L. Lopez, N.~E. Mavromatos, and D.~V. Nanopoulos, ``Precision
  tests of {CPT} symmetry and quantum mechanics in the neutral kaon system,''
  {\em Phys. Rev.} {\bf D53} (1996) 3846--3870,
\href{http://www.arXiv.org/abs/hep-ph/9505340}{{\tt hep-ph/9505340}}.

\bibitem{Banks:1984by}
T.~Banks, L.~Susskind, and M.~E. Peskin, ``Difficulties for the evolution of
  pure states into mixed states,'' {\em Nucl. Phys.} {\bf B244} (1984)
125;
W.~G. Unruh and R.~M. Wald, ``On evolution laws taking pure states to mixed
  states in quantum field theory,'' {\em Phys. Rev.} {\bf D52} (1995)
  2176--2182,
\href{http://www.arXiv.org/abs/hep-th/9503024}{{\tt hep-th/9503024}}.

\bibitem{Rajaraman:1982is}
R.~Rajaraman, {\em SOLITONS AND INSTANTONS. {A}N INTRODUCTION TO SOLITONS AND
  INSTANTONS IN QUANTUM FIELD THEORY}.
\newblock North-Holland, Amsterdam, 1982.


\bibitem{Kummer:1992bg}
W.~Kummer and D.~J. Schwarz, ``{General analytic solution of R**2 gravity with
  dynamical torsion in two-dimensions},'' {\em Phys. Rev.} {\bf D45} (1992)
3628--3635;
``Renormalization of {R}**2 gravity with dynamical
  torsion in d = 2,'' {\em Nucl. Phys.} {\bf B382} (1992)
171--186.

\bibitem{Brown:1988}
J.~Brown, {\em Lower Dimensional Gravity}.
\newblock World Scientific, 1988;
T.~Strobl, ``Gravity in two spacetime dimensions,''
  \href{http://www.arXiv.org/abs/hep-th/0011240}{{\tt hep-th/0011240}}.
Habilitation thesis;
S.~Nojiri and S.~D. Odintsov, ``Quantum dilatonic gravity in d = 2, 4 and 5
  dimensions,'' {\em Int. J. Mod. Phys.} {\bf A16} (2001) 1015--1108,
\href{http://arXiv.org/abs/hep-th/0009202}{{\tt hep-th/0009202}}.

\bibitem{Schaller:1994es}
P.~Schaller and T.~Strobl, ``Poisson structure induced (topological) field
  theories,'' {\em Mod. Phys. Lett.} {\bf A9} (1994) 3129--3136,
\href{http://arXiv.org/abs/hep-th/9405110}{{\tt hep-th/9405110}}.

\bibitem{Russo:1992yg}
J.~G. Russo and A.~A. Tseytlin, ``Scalar tensor quantum gravity in
  two-dimensions,'' {\em Nucl. Phys.} {\bf B382} (1992) 259--275,
\href{http://www.arXiv.org/abs/arXiv:hep-th/9201021}{{\tt
  arXiv:hep-th/9201021}};
S.~D. Odintsov and I.~L. Shapiro, ``One loop renormalization of two-dimensional
  induced quantum gravity,'' {\em Phys. Lett.} {\bf B263} (1991)
183--189.

\bibitem{Katanaev:2000kc}
M.~O. Katanaev, ``Effective action for scalar fields in two-dimensional
  gravity,'' {\em Annals Phys.} {\bf 296} (2002) 1--50,
\href{http://arXiv.org/abs/gr-qc/0101033}{{\tt gr-qc/0101033}}.

\bibitem{Grumiller:2001ea}
D.~Grumiller, {\em Quantum dilaton gravity in two dimensions with matter}.
\newblock PhD thesis, {T}echnische {U}niversit{\"a}t {W}ien, 2001.
\newblock
\href{http://www.arXiv.org/abs/gr-qc/0105078}{{\tt gr-qc/0105078}}.
\newblock

\bibitem{Katanaev:1993fu}
M.~O. Katanaev, ``All universal coverings of two-dimensional gravity with
  torsion,'' {\em J. Math. Phys.} {\bf 34} (1993)
700--736.

\bibitem{Kummer:1998zs}
W.~Kummer, H.~Liebl, and D.~V. Vassilevich, ``Integrating geometry in general
  2d dilaton gravity with matter,'' {\em Nucl. Phys.} {\bf B544} (1999)
  403--431,
\href{http://www.arXiv.org/abs/hep-th/9809168}{{\tt hep-th/9809168}}.

\bibitem{Liebl:1997ti}
H.~Liebl, D.~V. Vassilevich, and S.~Alexandrov, ``Hawking radiation and masses
  in generalized dilaton theories,'' {\em Class. Quant. Grav.} {\bf 14} (1997)
  889--904,
\href{http://www.arXiv.org/abs/arXiv:gr-qc/9605044}{{\tt arXiv:gr-qc/9605044}}.

\bibitem{Banks:1991mk}
T.~Banks and M.~O'Loughlin, ``Two-dimensional quantum gravity in {M}inkowski
  space,'' {\em Nucl. Phys.} {\bf B362} (1991)
649--664;
V.~P. Frolov, ``Two-dimensional black hole physics,'' {\em Phys. Rev.} {\bf
  D46} (1992)
5383--5394;
R.~B. Mann, ``Conservation laws and 2-d black holes in dilaton gravity,'' {\em
  Phys. Rev.} {\bf D47} (1993) 4438--4442,
\href{http://www.arXiv.org/abs/hep-th/9206044}{{\tt hep-th/9206044}}.

\bibitem{Grosse:1992vc}
H.~Grosse, W.~Kummer, P.~Presnajder, and D.~J. Schwarz, ``{Novel symmetry of
  nonEinsteinian gravity in two- dimensions},'' {\em J. Math. Phys.} {\bf 33}
  (1992) 3892--3900,
\href{http://www.arXiv.org/abs/hep-th/9205071}{{\tt hep-th/9205071}}.

\bibitem{Grumiller:2004wi}
D.~Grumiller and D.~Mayerhofer, ``On static solutions in 2d dilaton gravity
  with scalar matter,''
\href{http://www.arXiv.org/abs/gr-qc/0404013}{{\tt gr-qc/0404013}}.

\bibitem{Grumiller:2002dm}
D.~Grumiller, W.~Kummer, and D.~V. Vassilevich, ``Virtual black holes in
  generalized dilaton theories (and their special role in string gravity),''
  {\em European Phys. J.} {\bf C30} (2003) 135--143,
\href{http://arXiv.org/abs/hep-th/0208052}{{\tt hep-th/0208052}}.

\bibitem{Fischer:2001vz}
P.~Fischer, D.~Grumiller, W.~Kummer, and D.~V. Vassilevich, ``S-matrix for
  s-wave gravitational scattering,'' {\em Phys. Lett.} {\bf B521} (2001)
  357--363, \href{http://arXiv.org/abs/gr-qc/0105034}{{\tt gr-qc/0105034}}.
Erratum ibid. {\bf B532} (2002) 373.

\bibitem{Grumiller:2001rg}
D.~Grumiller, ``Virtual black hole phenomenology from 2d dilaton theories,''
  {\em Class. Quant. Grav.} {\bf 19} (2002) 997--1009,
\href{http://arXiv.org/abs/gr-qc/0111097}{{\tt gr-qc/0111097}}.

\bibitem{Grumiller:2003ez}
D.~Grumiller, ``Deformations of the {S}chwarzschild black hole,''
  \href{http://www.arXiv.org/abs/gr-qc/0311011}{{\tt gr-qc/0311011}}.
Invited talk at MG X.

\bibitem{Kuchar:1994zk}
K.~V. Kucha{\v{r}}, ``Geometrodynamics of {S}chwarzschild black holes,'' {\em
  Phys. Rev.} {\bf D50} (1994) 3961--3981,
\href{http://www.arXiv.org/abs/arXiv:gr-qc/9403003}{{\tt arXiv:gr-qc/9403003}}.

\bibitem{Argyres:1998qn}
P.~C. Argyres, S.~Dimopoulos, and J.~March-Russell, ``Black holes and
  sub-millimeter dimensions,'' {\em Phys. Lett.} {\bf B441} (1998) 96--104,
\href{http://www.arXiv.org/abs/hep-th/9808138}{{\tt hep-th/9808138}};
T.~Banks, M.~Dine, and A.~E. Nelson, ``Constraints on theories with large extra
  dimensions,'' {\em JHEP} {\bf 06} (1999) 014,
\href{http://www.arXiv.org/abs/hep-th/9903019}{{\tt hep-th/9903019}};
T.~Banks and W.~Fischler, ``A model for high energy scattering in quantum
  gravity,''
\href{http://www.arXiv.org/abs/hep-th/9906038}{{\tt hep-th/9906038}};
R.~Emparan, G.~T. Horowitz, and R.~C. Myers, ``Black holes radiate mainly on
  the brane,'' {\em Phys. Rev. Lett.} {\bf 85} (2000) 499--502,
\href{http://www.arXiv.org/abs/hep-th/0003118}{{\tt hep-th/0003118}}.

\bibitem{Dimopoulos:2001hw}
S.~B.~Giddings and S.~Thomas, ``High energy colliders as black hole factories: The end of short  distance physics,''
{\em Phys.\ Rev.} {\bf D65} (2002) 056010
\href{http://www.arXiv.org/abs/hep-ph/0106219}{{\tt hep-ph/0106219}};
S.~Dimopoulos and G.~Landsberg, ``{Black holes at the LHC},'' {\em Phys. Rev.
  Lett.} {\bf 87} (2001) 161602,
\href{http://arXiv.org/abs/hep-ph/0106295}{{\tt hep-ph/0106295}}.

\bibitem{Cavaglia:2002si}
M.~Cavaglia, ``{Black hole and brane production in TeV gravity: A review},''
  {\em Int. J. Mod. Phys.} {\bf A18} (2003) 1843--1882,
\href{http://www.arXiv.org/abs/hep-ph/0210296}{{\tt hep-ph/0210296}};
G.~Landsberg, ``{Black holes at future colliders and beyond: A review},''
\href{http://www.arXiv.org/abs/hep-ph/0211043}{{\tt hep-ph/0211043}};
``Black holes at future colliders and in cosmic rays,''
\href{http://www.arXiv.org/abs/hep-ex/0310034}{{\tt hep-ex/0310034}}.

\bibitem{Arkani-Hamed:1998rs}
N.~Arkani-Hamed, S.~Dimopoulos, and G.~R. Dvali, ``The hierarchy problem and
  new dimensions at a millimeter,'' {\em Phys. Lett.} {\bf B429} (1998)
  263--272,
\href{http://www.arXiv.org/abs/hep-ph/9803315}{{\tt hep-ph/9803315}};
I.~Antoniadis, N.~Arkani-Hamed, S.~Dimopoulos, and G.~R. Dvali, ``{New
  dimensions at a millimeter to a Fermi and superstrings at a TeV},'' {\em
  Phys. Lett.} {\bf B436} (1998) 257--263,
\href{http://www.arXiv.org/abs/hep-ph/9804398}{{\tt hep-ph/9804398}};
N.~Arkani-Hamed, S.~Dimopoulos, and G.~R. Dvali, ``{Phenomenology, astrophysics
  and cosmology of theories with sub-millimeter dimensions and TeV scale
  quantum gravity},'' {\em Phys. Rev.} {\bf D59} (1999) 086004,
\href{http://www.arXiv.org/abs/hep-ph/9807344}{{\tt hep-ph/9807344}}.

\bibitem{Zeldovich:1976vq}
Y.~B.~Zeldovich, ``A New Type Of Radioactive Decay: Gravitational Annihilation Of Baryons,''
{\em Phys.\ Lett.} {\bf A59}, 254 (1976);
``A Novel Type Of Radioactive Decay: Gravitational Baryon Annihilation,'' (In Russian), {\em Zh.\ Eksp.\ Teor.\ Fiz.}  {\bf 72}, 18 (1977);
S.~W.~Hawking, D.~N.~Page and C.~N.~Pope,
``The Propagation Of Particles In Space-Time Foam,''
{\em Phys.\ Lett.} {\bf B86}, 175 (1979);
D.~N.~Page,
``Particle Transmutations In Quantum Gravity,''
{\em Phys.\ Lett.} {\bf B95}, 244 (1980).

\bibitem{Adams:2000za}
F.~C.~Adams, G.~L.~Kane, M.~Mbonye and M.~J.~Perry, ``Proton decay, black holes, and large extra dimensions,''
{\em Int.\ J.\ Mod.\ Phys.} {\bf A16} (2001) 2399,
\href{http://www.arXiv.org/abs/hep-ph/0009154}{{\tt hep-ph/0009154}}.

\bibitem{Kobakhidze:2001yk}
A.~B.~Kobakhidze, ``Proton stability in TeV-scale GUTs,''
{\em Phys.\ Lett.} {\bf B514} (2001) 131,
\href{http://www.arXiv.org/abs/hep-ph/0102323}{{\tt hep-ph/0102323}}.

\bibitem{Uehara:2002cj}
Y.~Uehara, ``Virtual black holes at linear colliders,''
\href{http://arXiv.org/abs/hep-ph/0205068}{{\tt hep-ph/0205068}}.

\bibitem{'tHooft:1996tq}
G.~'t~Hooft, ``{The scattering matrix approach for the quantum black hole: An
  overview},'' {\em Int. J. Mod. Phys.} {\bf A11} (1996) 4623--4688,
\href{http://www.arXiv.org/abs/gr-qc/9607022}{{\tt gr-qc/9607022}}.

\bibitem{fischervertices}
P.~Fischer, ``Vertices in spherically reduced quantum gravity,'' Master's
  thesis, Vienna University of Technology, 2001;
D.~Grumiller, 2001.
\newblock Documented {\em Mathematica} notepad available from
  \href{http://www.teilchen.at/grumiller/projects/myself/thesis/s4.nb}{{\tt
  http://www.teilchen.at/grumiller/projects/myself/thesis/s4.nb}}.

\bibitem{Grumiller:2003dh}
D.~Grumiller, ``Three functions in dilaton gravity: The good, the bad and the
  muggy,'' in {\em Proceedings of International Workshop on Mathematical
  Theories and their Applications}, S.~Moskaliuk, ed., pp.~56--96, TIMPANI.
\newblock Cernivtsi, Ukraine, 2004.
\newblock
\href{http://www.arXiv.org/abs/hep-th/0305073}{{\tt hep-th/0305073}}.
\newblock

\bibitem{Grumiller:2003sk}
D.~Grumiller and W.~Kummer, ``How to approach quantum gravity: Background
  independence in 1+1 dimensions,'' in {\em Symmetries beyond the standard model},  N.~Mankoc Borstnik, H.~B. Nielsen, C.~D. Froggatt and D.~Lukman, eds., pp.~184--196 \newblock  Portoroz, Slovenia, 2003.
  \href{http://www.arXiv.org/abs/gr-qc/0310068}{{\tt gr-qc/0310068}}.

\bibitem{Ashtekar:1991hf}
A.~Ashtekar, ``Lectures on nonperturbative canonical gravity,''. Singapore,
  Singapore: World Scientific (1991) 334 p. (Advanced series in astrophysics
  and cosmology, 6).

\bibitem{Balasin:1992nm}
H.~Balasin, W.~Kummer, O.~Piguet, and M.~Schweda, ``{On the regularization of
  the mass zero 2-D propagator},'' {\em Phys. Lett.} {\bf B287} (1992)
138--144.

\bibitem{Moretti:2002mp}
V.~Moretti and N.~Pinamonti, ``{Aspects of hidden and manifest SL(2,R) symmetry
  in 2D near- horizon black-hole backgrounds},'' {\em Nucl. Phys.} {\bf B647}
  (2002) 131--152,
\href{http://www.arXiv.org/abs/gr-qc/0207072}{{\tt gr-qc/0207072}}.

\bibitem{Wald:1999vt}
R.~M. Wald, ``The thermodynamics of black holes,'' {\em Living Rev. Rel.} {\bf
  4} (2001) 6,
\href{http://www.arXiv.org/abs/gr-qc/9912119}{{\tt gr-qc/9912119}}.

\bibitem{Gibbons:1994cg}
G.~W. Gibbons and S.~W. Hawking, eds., {\em Euclidean quantum gravity}.
\newblock Singapore: World Scientific, 1993.

\bibitem{Kummer:1999zy}
W.~Kummer and D.~V. Vassilevich, ``{Hawking radiation from dilaton gravity in
  (1+1) dimensions: A pedagogical review},'' {\em Annalen Phys.} {\bf 8} (1999)
  801--827,
\href{http://arXiv.org/abs/gr-qc/9907041}{{\tt gr-qc/9907041}}.

\bibitem{Bekenstein:1973ur}
J.~D. Bekenstein, ``Black holes and entropy,'' {\em Phys. Rev.} {\bf D7} (1973)
2333--2346;
S.~W. Hawking, ``Particle creation by black holes,'' {\em Commun. Math. Phys.}
  {\bf 43} (1975)
199--220.

\bibitem{Gegenberg:1995pv}
J.~Gegenberg, G.~Kunstatter, and D.~Louis-Martinez, ``Observables for
  two-dimensional black holes,'' {\em Phys. Rev.} {\bf D51} (1995) 1781--1786,
\href{http://arXiv.org/abs/gr-qc/9408015}{{\tt gr-qc/9408015}}.

\bibitem{Youm:1997hw}
D.~Youm, ``Black holes and solitons in string theory,'' {\em Phys. Rept.} {\bf
  316} (1999) 1--232,
\href{http://www.arXiv.org/abs/hep-th/9710046}{{\tt hep-th/9710046}};
R.~D'Auria and P.~Fre, ``{BPS black holes in supergravity: Duality groups,
  p-branes, central charges and the entropy},''
\href{http://www.arXiv.org/abs/hep-th/9812160}{{\tt hep-th/9812160}};
T.~Mohaupt, ``Black hole entropy, special geometry and strings,'' {\em Fortsch.
  Phys.} {\bf 49} (2001) 3--161,
\href{http://www.arXiv.org/abs/hep-th/0007195}{{\tt hep-th/0007195}}.

\bibitem{Ashtekar:2000eq}
A.~Ashtekar, J.~C. Baez, and K.~Krasnov, ``Quantum geometry of isolated
  horizons and black hole entropy,'' {\em Adv. Theor. Math. Phys.} {\bf 4}
  (2000) 1--94,
\href{http://www.arXiv.org/abs/gr-qc/0005126}{{\tt gr-qc/0005126}}.

\bibitem{Carlip:2004mn}
S.~Carlip, ``Horizon constraints and black hole entropy,''
\href{http://www.arXiv.org/abs/hep-th/0408123}{{\tt hep-th/0408123}}.

\bibitem{Hod:1998vk}
S.~Hod, ``Bohr's correspondence principle and the area spectrum of quantum
  black holes,'' {\em Phys. Rev. Lett.} {\bf 81} (1998) 4293,
\href{http://www.arXiv.org/abs/gr-qc/9812002}{{\tt gr-qc/9812002}}.

\bibitem{Kettner:2004aw}
J.~Kettner, G.~Kunstatter, and A.~J.~M. Medved, ``Quasinormal modes for single
  horizon black holes in generic 2-d dilaton gravity,''
\href{http://www.arXiv.org/abs/gr-qc/0408042}{{\tt gr-qc/0408042}}.

\bibitem{Cadoni:1998sg}
M.~Cadoni and S.~Mignemi, ``Entropy of 2d black holes from counting
  microstates,'' {\em Phys. Rev.} {\bf D59} (1999) 081501,
\href{http://www.arXiv.org/abs/hep-th/9810251}{{\tt hep-th/9810251}}.

\bibitem{Carlip:1999cy}
S.~Carlip, ``Entropy from conformal field theory at killing horizons,'' {\em
  Class. Quant. Grav.} {\bf 16} (1999) 3327--3348,
\href{http://www.arXiv.org/abs/gr-qc/9906126}{{\tt gr-qc/9906126}}.

\bibitem{Carlip:2002be}
S.~Carlip, ``Near-horizon conformal symmetry and black hole entropy,'' {\em
  Phys. Rev. Lett.} {\bf 88} (2002) 241301,
\href{http://www.arXiv.org/abs/gr-qc/0203001}{{\tt gr-qc/0203001}}.

\bibitem{Carlip:1998wz}
S.~Carlip, ``Black hole entropy from conformal field theory in any dimension,''
  {\em Phys. Rev. Lett.} {\bf 82} (1999) 2828--2831,
\href{http://www.arXiv.org/abs/hep-th/9812013}{{\tt hep-th/9812013}};
S.~N. Solodukhin, ``Conformal description of horizon's states,'' {\em Phys.
  Lett.} {\bf B454} (1999) 213--222,
\href{http://www.arXiv.org/abs/hep-th/9812056}{{\tt hep-th/9812056}}.

\bibitem{Schwarzschild:1916uq}
K.~Schwarzschild, ``On the gravitational field of a mass point according to
  {E}instein's theory,'' {\em Sitzungsber. Preuss. Akad. Wiss. Berlin (Math.
  Phys. )} {\bf 1916} (1916) 189--196,
\href{http://www.arXiv.org/abs/arXiv:physics/9905030}{{\tt
  arXiv:physics/9905030}}.

\bibitem{Mandal:1991tz}
G.~Mandal, A.~M. Sengupta, and S.~R. Wadia, ``Classical solutions of
  two-dimensional string theory,'' {\em Mod. Phys. Lett.} {\bf A6} (1991)
1685--1692;
S.~Elitzur, A.~Forge, and E.~Rabinovici, ``Some global aspects of string
  compactifications,'' {\em Nucl. Phys.} {\bf B359} (1991)
581--610;
E.~Witten, ``On string theory and black holes,'' {\em Phys. Rev.} {\bf D44}
  (1991)
314--324;
R.~Dijkgraaf, H.~Verlinde, and E.~Verlinde, ``String propagation in a black
  hole geometry,'' {\em Nucl. Phys.} {\bf B371} (1992)
269--314;
C.~G. Callan, Jr., S.~B. Giddings, J.~A. Harvey, and A.~Strominger,
  ``Evanescent black holes,'' {\em Phys. Rev.} {\bf D45} (1992) 1005--1009,
\href{http://www.arXiv.org/abs/hep-th/9111056}{{\tt hep-th/9111056}}.

\bibitem{Teitelboim:1983ux}
C.~Teitelboim, ``Gravitation and {H}amiltonian structure in two space-time
  dimensions,'' {\em Phys. Lett.} {\bf B126} (1983)
41;
R.~Jackiw, ``Lower dimensional gravity,'' {\em Nucl. Phys.} {\bf B252} (1985)
343--356.

\bibitem{Katanaev:1997ni}
M.~O. Katanaev, W.~Kummer, and H.~Liebl, ``On the completeness of the black
  hole singularity in 2d dilaton theories,'' {\em Nucl. Phys.} {\bf B486}
  (1997) 353--370,
\href{http://www.arXiv.org/abs/gr-qc/9602040}{{\tt gr-qc/9602040}}.

\bibitem{Hawking:1982dh}
S.~W. Hawking and D.~N. Page, ``Thermodynamics of black holes in anti-de
  {S}itter space,'' {\em Commun. Math. Phys.} {\bf 87} (1983)
577.

\bibitem{Guralnik:2003we}
G.~Guralnik, A.~Iorio, R.~Jackiw, and S.~Y. Pi, ``{Dimensionally reduced
  gravitational Chern-Simons term and its kink},'' {\em Ann. Phys.} {\bf 308}
  (2003) 222--236,
\href{http://www.arXiv.org/abs/hep-th/0305117}{{\tt hep-th/0305117}};
D.~Grumiller and W.~Kummer, ``{The classical solutions of the dimensionally
  reduced gravitational Chern-Simons theory},'' {\em Ann. Phys.} {\bf 308}
  (2003) 211--221,
\href{http://www.arXiv.org/abs/hep-th/0306036}{{\tt hep-th/0306036}};
L.~Bergamin, D.~Grumiller, A.~Iorio, and C.~Nu{\~n}ez, ``Chemistry of Chern-Simons Supergravity: reduction to a BPS kink, oxidation to M-theory and thermodynamical aspects,'' \href{http://www.arXiv.org/abs/hep-th/0409273}{{\tt hep-th/0409273}}.

\bibitem{Aharony:1999ti}
O.~Aharony, S.~S. Gubser, J.~M. Maldacena, H.~Ooguri, and Y.~Oz, ``{Large N
  field theories, string theory and gravity},'' {\em Phys. Rept.} {\bf 323}
  (2000) 183--386,
\href{http://www.arXiv.org/abs/hep-th/9905111}{{\tt hep-th/9905111}}.

\bibitem{Brown:1994gs}
J.~D. Brown, J.~Creighton, and R.~B. Mann, ``{Temperature, energy and heat
  capacity of asymptotically anti-de Sitter black holes},'' {\em Phys. Rev.}
  {\bf D50} (1994) 6394--6403,
\href{http://www.arXiv.org/abs/gr-qc/9405007}{{\tt gr-qc/9405007}}.

\bibitem{Zaslavsky:1996dg}
O.~B. Zaslavsky, ``Quantum corrections to temperature and mass of 1+1 dilatonic
  black holes and the trace anomaly,'' {\em Phys. Lett.} {\bf B375} (1996)
43--46.

\bibitem{Grumiller:2003mc}
D.~Grumiller, W.~Kummer, and D.~V. Vassilevich, ``Positive specific heat of the
  quantum corrected dilaton black hole,'' {\em JHEP} {\bf 07} (2003) 009,
\href{http://www.arXiv.org/abs/hep-th/0305036}{{\tt hep-th/0305036}}.

\bibitem{Hawking:1971tu}
S.~W. Hawking, ``Gravitational radiation from colliding black holes,'' {\em
  Phys. Rev. Lett.} {\bf 26} (1971)
1344--1346;
S.~W. Hawking, ``Black holes in general relativity,'' {\em Commun. Math. Phys.}
  {\bf 25} (1972)
152--166.

\bibitem{Corless:1996}
R.~M. Corless, G.~H. Gonnet, D.~E.~G. Hare, D.~J. Jeffrey, and D.~E. Knuth,
  ``{On the Lambert W Function},'' {\em Adv. Comp. Math.} {\bf 5} (1996)
  329--359. More information on the Lambert $W$ function is available at the
  webpage
  \href{http://kong.apmaths.uwo.ca/$\sim$rcorless/frames/PAPERS/LambertW/}{{\tt
  http://kong.apmaths.uwo.ca/$\sim$rcorless/frames/PAPERS/LambertW/}}.

\bibitem{Flanagan:2004bz}
E.~E. Flanagan, ``The conformal frame freedom in theories of gravitation,''
  {\em Class. Quant. Grav.} {\bf 21} (2004) 3817,
\href{http://www.arXiv.org/abs/gr-qc/0403063}{{\tt gr-qc/0403063}}.

\bibitem{Mann:1998hm}
R.~B. Mann and S.~N. Solodukhin, ``Universality of quantum entropy for extreme
  black holes,'' {\em Nucl. Phys.} {\bf B523} (1998) 293--307,
\href{http://www.arXiv.org/abs/hep-th/9709064}{{\tt hep-th/9709064}};
R.~K. Kaul and P.~Majumdar, ``{Logarithmic correction to the Bekenstein-Hawking
  entropy},'' {\em Phys. Rev. Lett.} {\bf 84} (2000) 5255--5257,
\href{http://www.arXiv.org/abs/gr-qc/0002040}{{\tt gr-qc/0002040}};
S.~Carlip, ``{Logarithmic corrections to black hole entropy from the Cardy
  formula},'' {\em Class. Quant. Grav.} {\bf 17} (2000) 4175--4186,
\href{http://www.arXiv.org/abs/gr-qc/0005017}{{\tt gr-qc/0005017}}.

\bibitem{Ghosh:2004rq}
A.~Ghosh and P.~Mitra, ``A bound on the log correction to the black hole area
  law,''
\href{http://www.arXiv.org/abs/gr-qc/0401070}{{\tt gr-qc/0401070}}.

\bibitem{Fiola:1994ir}
T.~M. Fiola, J.~Preskill, A.~Strominger, and S.~P. Trivedi, ``Black hole
  thermodynamics and information loss in two- dimensions,'' {\em Phys. Rev.}
  {\bf D50} (1994) 3987--4014,
\href{http://www.arXiv.org/abs/hep-th/9403137}{{\tt hep-th/9403137}};
R.~C. Myers, ``Black hole entropy in two-dimensions,'' {\em Phys. Rev.} {\bf
  D50} (1994) 6412--6421,
\href{http://www.arXiv.org/abs/hep-th/9405162}{{\tt hep-th/9405162}};
J.~D. Hayward, ``{Entropy in the RST model},'' {\em Phys. Rev.} {\bf D52}
  (1995) 2239--2244,
\href{http://www.arXiv.org/abs/gr-qc/9412065}{{\tt gr-qc/9412065}}.

\bibitem{Giddings:2001pt}
S.~B.~Giddings and M.~Lippert, ``Precursors, black holes, and a locality bound,''
{\em Phys.\ Rev.} {\bf D65} (2002) 024006
\href{http://www.arXiv.org/abs/hep-th/0103231}{{\tt hep-th/0103231}};
``The information paradox and the locality bound,''
{\em Phys.\ Rev.} {\bf D69} (2004) 124019
\href{http://www.arXiv.org/abs/hep-th/0402073}{{\tt hep-th/0402073}}.

\bibitem{Bergamin:2004us}
L.~Bergamin, D.~Grumiller, and W.~Kummer, ``Quantization of 2d dilaton
  supergravity with matter,'' {\em JHEP} {\bf 05} (2004) 060,
\href{http://www.arXiv.org/abs/hep-th/0404004}{{\tt hep-th/0404004}}.

\bibitem{Bojowald:2004}
M.~Bojowald and R.~Swiderski, ``Spherically symmetric quantum horizons,'' in preparation.

\end{thebibliography}
\providecommand{\href}[2]{#2}\begingroup\raggedright\endgroup

\end{document}